\pdfoutput=1
\documentclass[11pt]{article}
\usepackage{acl}
\usepackage{times}
\usepackage{amsmath}
\usepackage{latexsym}
\usepackage[T1]{fontenc}
\usepackage[utf8]{inputenc}
\usepackage{microtype}
\usepackage{inconsolata}
\usepackage{multirow}
\usepackage{booktabs}
\usepackage{graphics}
\usepackage{subcaption}
\usepackage[export]{adjustbox}
\usepackage{fontawesome}
\usepackage{pgfplots}
\pgfplotsset{compat=1.18}
\usepackage{tikz}
\usetikzlibrary{calc,shapes,arrows,shadows,shapes.callouts,shapes.arrows,chains,positioning,trees, matrix}
\usepackage{fancyvrb}

\title{Entity Retrieval for Answering Entity-Centric Questions}

\author{Hassan S. Shavarani \\
  School of Computing Science \\
  Simon Fraser University \\
  BC, Canada \\
  \texttt{sshavara@sfu.ca} \\\And
  Anoop Sarkar \\
  School of Computing Science \\
  Simon Fraser University\\
  BC, Canada \\
  \texttt{anoop@sfu.ca}
}

\begin{document}
\maketitle
\begin{abstract}
 The similarity between the question and indexed documents is a crucial factor in document retrieval for retrieval-augmented question answering. Although this is typically the only method for obtaining the relevant documents, it is not the sole approach when dealing with entity-centric questions. In this study, we propose \textit{Entity Retrieval}, a novel retrieval method which rather than relying on question-document similarity, depends on the salient entities within the question to identify the retrieval documents. We conduct an in-depth analysis of the performance of both dense and sparse retrieval methods in comparison to \textit{Entity Retrieval}. Our findings reveal that our method not only leads to more accurate answers to entity-centric questions but also operates more efficiently.

\faGithub~\href{https://github.com/shavarani/EntityRetrieval}{https://github.com/shavarani/EntityRetrieval}
\end{abstract}
\section{Introduction}

Information retrieval has significantly enhanced the factual reliability of large language model (LLM) generated responses \cite{2021.findings-emnlp.320} in question answering \citep{arXiv:2101.00774, 2023.acl-long.808}. This improvement is particularly evident in Retrieval-Augmented Generation \cite[RAG;][]{RAG, 2021.eacl-main.74, EMDR2}, which typically employs the Retriever-Reader architecture \citep{P17-1171}. RAG retrievers can be sparse \cite{arXiv:2302.12813}, dense \cite{2020.emnlp-main.550}, or hybrid \cite{2022.naacl-main.194}, while the readers are usually generative language models\footnote{The readers in the original architecture were designed to extract answer spans rather than generate answers.} such as BART \citep{2020.acl-main.703}, T5 \citep{T5}, or GPT-4 \citep{GPT4} that generate answers based on the documents identified by the retriever. Recent RAG methodologies leverage the in-context learning capabilities of LLMs to incorporate retrieved documents into the prompt \citep{arXiv:2301.12652, arXiv:2302.12813, arXiv:2305.14002}.

Entity-centric questions seek concise factual answers about the real world, typically in the form of single words or short phrases. These answers often reference or directly stem from a knowledge base entity \citep{IEEE_7917964}, and Retrieval-Augmentation enhances LLM performance in answering such questions, particularly for rare entities that appear infrequently in LLM training and fine-tuning data \citep{LLMLearnLongTail?}.

But is there a correlation between the quality of the retrieved documents and the generated response quality? \citet{2021.emnlp-main.496} found that dense retrievers retrieve less relevant documents for answering entity-centric questions than simpler sparse retrievers. Additionally, \citet{arXiv:2401.14887} show that the presence of irrelevant documents leads to worse answers. These findings underscore the crucial role of the retrieval module, particularly for entity-centric questions.

\begin{figure*}[t]
    \centering
    \begin{subfigure}{1.\columnwidth}
        \begin{center}
            \begin{tikzpicture}
                \fill[red!30]   (0.4,3.5) -- (0.4,2) .. controls (0.4,1.8) and (1.3,1.6) .. (2.4,1.6) .. controls (3.5,1.6) and (4.4,1.8) .. (4.4,2) -- (4.4,3.5)(0.4,3.5) .. controls (0.4,3.3) and (1.3,3.1) .. (2.4,3.1) .. controls (3.5,3.1) and (4.4,3.3) .. (4.4,3.5) .. controls (4.4,3.7) and (3.5,3.9) .. (2.4,3.9) .. controls (1.3,3.9) and (0.4,3.7) .. (0.4,3.5) -- cycle ;
                \draw   (0.4,3.5) -- (0.4,2) .. controls (0.4,1.8) and (1.3,1.6) .. (2.4,1.6) .. controls (3.5,1.6) and (4.4,1.8) .. (4.4,2) -- (4.4,3.5)(0.4,3.5) .. controls (0.4,3.3) and (1.3,3.1) .. (2.4,3.1) .. controls (3.5,3.1) and (4.4,3.3) .. (4.4,3.5) .. controls (4.4,3.7) and (3.5,3.9) .. (2.4,3.9) .. controls (1.3,3.9) and (0.4,3.7) .. (0.4,3.5) -- cycle ;
                \node at (2.4,3.5) {21 Million Passages};
                \node at (2.4,2.35) {\includegraphics[width=1.5cm]{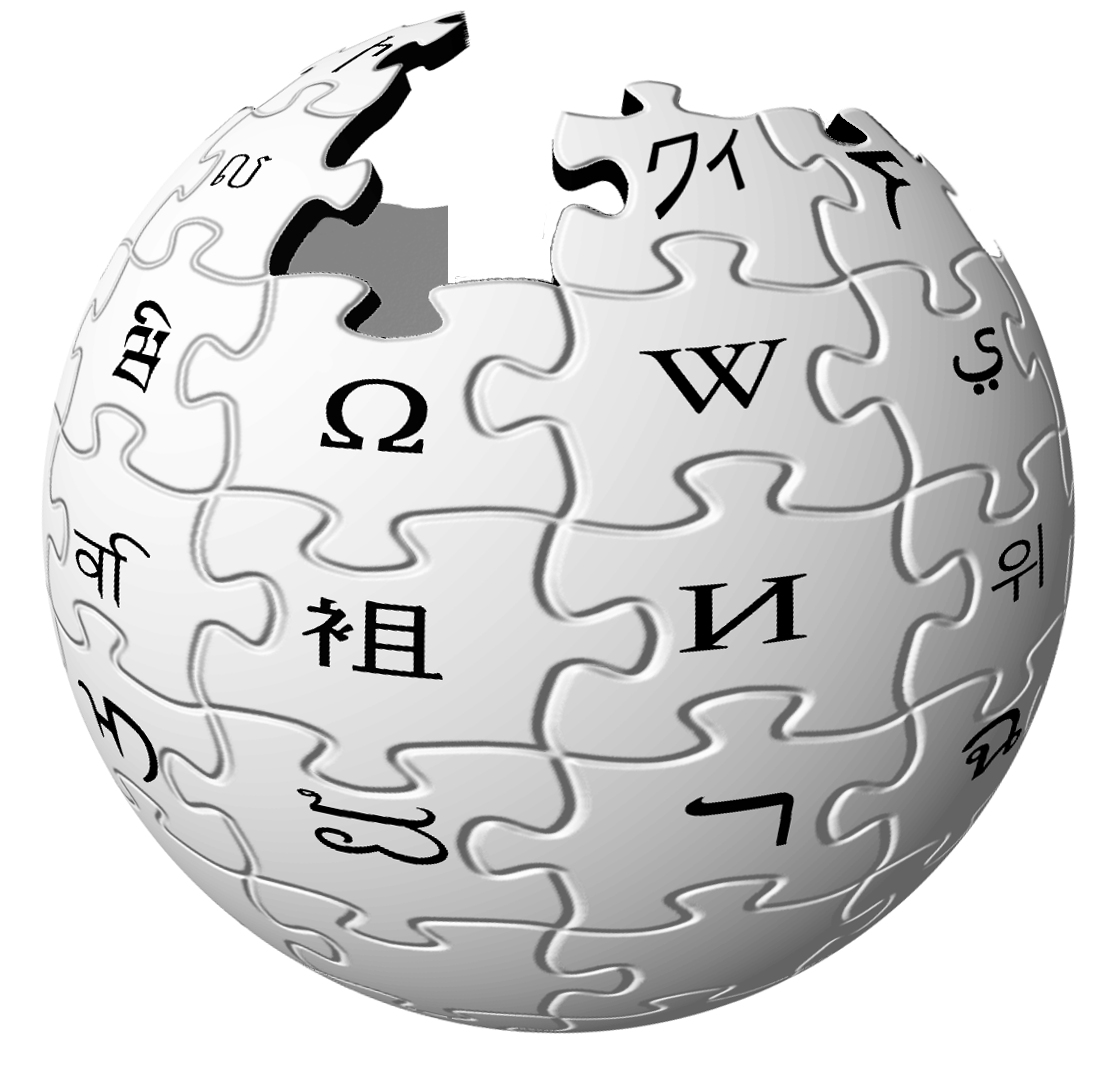}};

                \draw[->] (2.4,1.6) -- (2.4,1.3);
                \draw[rounded corners=3mm, draw=gray, line width=0.8pt, fill=cyan!40] (1.15,0) rectangle ++(2.5, 1.3);
                
                \node at (2.4,0.6) {\begin{tabular}{c}Dense/Sparse\\Retrieval\end{tabular}};
                \node at (0.2,0.7) {\textbf{Q}};
                \draw[->] (0.4,0.7) -- (1.15,0.7);

                \draw[->] (3.65,0.7) -- (4.4,0.7);
                \draw[->] (0.65, 0.7) -- (0.65, -0.1) -- (4.9, -0.1) -- (4.9, 0.05);
                \draw[rounded corners=3mm, draw=gray, line width=0.8pt, fill=magenta!40] (4.4,0.05) rectangle ++(1.3, 1.3);
                \node at (5.05,0.65) {LLM};
        
                \draw[->] (5.7,0.7) -- (6.45,0.7);
                \node at (6.65,0.7) {\textbf{A}};
            \end{tikzpicture}
        \end{center}
        \caption{Retrieval-Augmented QA with Dense/Sparse Retrieval}
        \label{fig:dpr_process}
    \end{subfigure}
    \begin{subfigure}{1.\columnwidth}
        \begin{center}
            \begin{tikzpicture}
                \node at (0,0.95) {\textbf{Q}};
                
                \draw[->] (0.2,0.95) -- (0.7,0.95);
                
                \draw[rounded corners=3mm, draw=gray, line width=0.8pt, fill=cyan!40] (0.7,0.25) rectangle ++(1.45, 1.3);
                \node at (1.45,0.85) {\begin{tabular}{c}Identify\\Entities\end{tabular}};
                
                \draw[->] (2.15,0.95) -- (2.65,0.95);
                
                \draw[gray, line width=0.8pt, fill=red!40] (3.025,0) -- (3.9,0) -- (3.9,1.75) -- (2.65,1.75) -- (2.65,0.375) -- cycle -- (3.025,0); 
                \draw[gray, line width=0.8pt] (2.65,0.375) -- (2.95,0.3) -- (3.025,0);
                \node at (3.27,0.96) {\begin{tabular}{c}Lookup\\Entity\\Articles\end{tabular}};
                
                \draw[->] (3.9,0.95) -- (4.4,0.95);
                
                \draw[rounded corners=3mm, draw=gray, line width=0.8pt, fill=green!40] (4.4,0.25) rectangle ++(2, 1.3);
                \node at (5.4,0.85) {\begin{tabular}{c}Fetch First\\$W$ Words\end{tabular}};
            
                \draw[->] (5.3,1.55) -- (5.3,2);
                
                \draw[->] (0, 1.2) -- (0, 2.75) -- (4.65, 2.75);
            
                \draw[->] (5.95,2.75) -- (6.45,2.75);
                \node at (6.65,2.75) {\textbf{A}};
            
                \draw[rounded corners=3mm, draw=gray, line width=0.8pt, fill=magenta!40] (4.65,2) rectangle ++(1.3, 1.3);
                \node at (5.3,2.65) {LLM};
            \end{tikzpicture}
        \end{center}
        \caption{Retrieval-Augmented QA with \textit{Entity Retrieval}}
        \label{fig:entity_retrieval_process}
    \end{subfigure}
    \caption{\textit{Entity Retrieval} simplifies the process of obtaining augmentation documents by replacing the need to search through large indexed passages with a straightforward lookup. For \textbf{Q}: \texttt{What is the capital of Seine-Saint-Denis?} \textit{Entity Retrieval} considers the first few sentences of \texttt{Seine-Saint-Denis} Wikipedia article which states ``\texttt{Its prefecture is \textbf{Bobigny}.}'' and returns \textbf{A = Bobigny} where the other retrieval methods return \textbf{A = Saint-Denis} or \textbf{A = Paris}.}
    \label{fig:dpr_to_entity_retrieval_comparison}
    \vspace{-0.3cm}
\end{figure*}
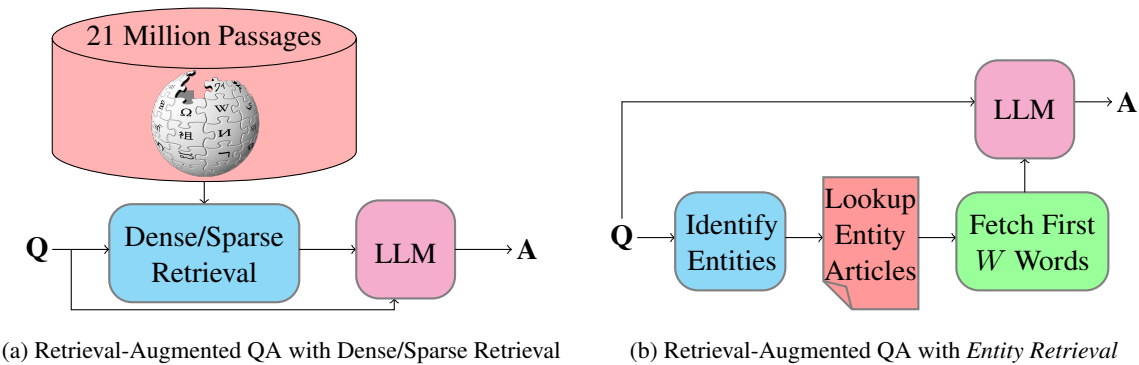

In this paper, we propose \textit{Entity Retrieval} (Figure \ref{fig:entity_retrieval_process}), which uses salient entities in the question to lookup knowledge base (e.g., Wikipedia) articles that correspond to each entity. Each article is truncated to the first $W$ words to form a document set that augments the question passed to the LLM.

Our contributions are as follows: (1) we propose \textit{Entity Retrieval}, a novel method of acquiring augmentation documents using salient entities in the questions, (2) we compare the retrieval performance quality of several retrieval techniques (both dense and sparse) to \textit{Entity Retrieval} for questions within two entity-centric question answering datasets, (3) we study the Retrieval-Augmentation quality of the compared techniques and \textit{Entity Retrieval}, using salient entity annotations of the questions, and (4) we examine the application of a recent state-of-the-art entity linking method for \textit{Entity Retrieval} in the absence of entity annotations in entity-centric questions.

\section{Retrieval for Retrieval-Augmentation}\label{sec:retrieval_methods}
Retrieval-Augmentation \cite{RAG} can be employed as a method of converting Closed-book question answering\footnote{Closed-book QA focuses on answering questions without additional context during inference.} \cite{2020.emnlp-main.437} into extractive question answering \cite{A00-1041, D16-1264}, where the answers can be directly extracted from the retrieved documents. Despite the abundance of effective retrieval techniques for Retrieval-Augmented Question Answering in existing literature \citep[][\textit{inter alia.}]{LTRe, RepBERT, 2021.acl-short.123, 2022.findings-emnlp.19, Contriever, 2022.naacl-main.272, 2022.emnlp-main.669}, this section will concentrate on a select few methods\footnote{We selected the methods supported by \url{pyserini.io} for the similarity between the underlying modules, minimizing discrepancies across different implementations.} utilized to study answering entity-centric questions in this paper.

\textbf{BM25} \citep{BM25Original, BM25} is a probabilistic retrieval method that ranks documents based on the frequency of query terms appearing in each document, adjusted by the length of the document and overall term frequency in the collection. It operates in the sparse vector space, relying on precomputed term frequencies and inverse document frequencies to retrieve documents based on keyword matching.

\textbf{DPR} \citep[Dense Passage Retrieval;][]{2020.emnlp-main.550} leverages a bi-encoder architecture, wherein the initial encoder processes the question and the subsequent encoder handles the passages to be retrieved. The similarity scores between the two encoded representations are computed using a dot product. Typically, the encoded representations of the second encoder are fixed and indexed in FAISS \cite{FAISS}, while the first encoder is optimized to maximize the dot-product scores based on positive and negative examples.

\textbf{ANCE} \citep{ANCE} is another dense retrieval technique similar to DPR\footnote{We have also implemented {DKRR} \citep{DKRR}, however, due to its significantly poorer performance compared to other methods, we exclude it from our analysis.}. It employs one encoder to transform both the questions and passages into dense representations. The key distinction from DPR is that ANCE uses hard negatives generated by periodically updating the passage embeddings during training, which helps the model learn more discriminative features, thereby enhancing retrieval performance over time.

\section{Entity Retrieval for Question Answering}\label{sec:el_for_qa}
While quite powerful, most Retrieval-Augmented systems are notably time and resource-intensive, necessitating the storage of extensive lookup indices and the need to attend to all retrieved documents to generate the response (see Section \ref{sec:raqa_efficiency_analysis}). This attribute renders such methods less desirable, particularly given the drive to run LLMs locally and on mobile phones \cite{arXiv:2312.11514}.

Entity recognition has been an integral component of statistical question answering systems \cite[\textit{inter alia}]{W16-0104}. Additionally, the extensively studied field of Knowledge Base Question Answering \citep[\textit{inter alia}]{KBQA} has underscored the significance of entity information from knowledge bases in question answering \cite{2023.paclic-1.63}.
A traditional neural question answering pipeline may contain entity detection, entity linking, relation prediction, and evidence integration \citep{N18-2047, arXiv:2001.11985}, where entity detection can employ LSTM-based \cite{LSTM} or BERT-based \cite{N19-1423} encoders.
Inspired by this body of work, we investigate the relevance of retrieval based on entity information as an alternative strategy to the proposed retrieval methods of Section \ref{sec:retrieval_methods}, especially for answering entity-centric questions with LLMs.

\begin{figure}[t]
    \centering
    \includegraphics [width=0.47\textwidth]{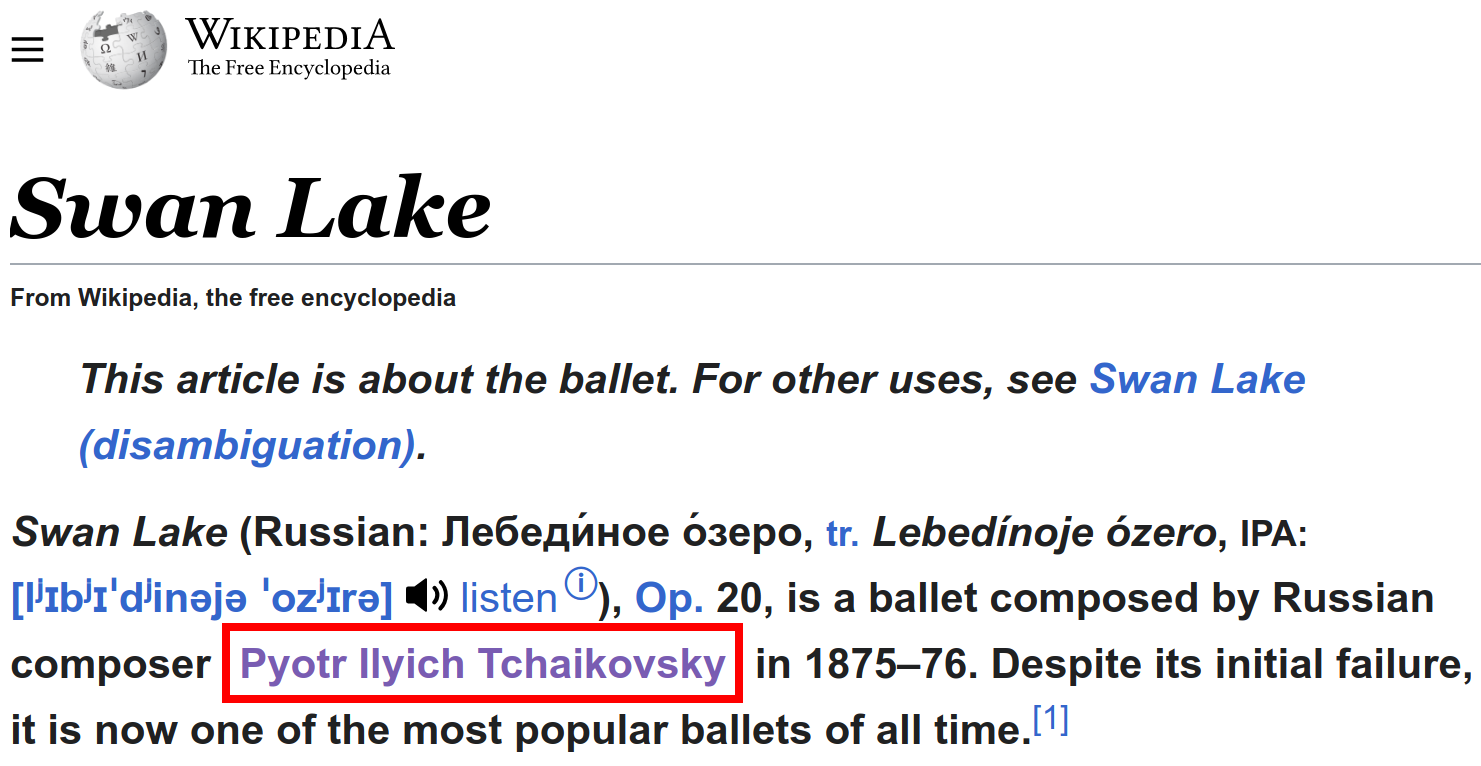}
    \caption{The first paragraph of the Wikipedia article typically provides an informative summary for the entity. For example, the first paragraph of \texttt{Swan Lake} Wikipedia article contains the answer to ``\texttt{Who is the composer of The Swan Lake ballet?}''}
    \label{fig:entiy_response_example}
\end{figure}

Our proposed method, \textit{Entity Retrieval}, leverages the salient entities within the questions to identify and retrieve their corresponding knowledge base articles. We will then truncate these articles to the first $W$ words\footnote{The first sentences of Wikipedia articles have been proven informative for document classification \citep{2020.lrec-1.150} as well as question answering \citep{D18-1241}.} to form the list of the documents augmenting entity-centric questions when prompting LLMs. Figure \ref{fig:dpr_to_entity_retrieval_comparison} presents a schematic comparison between \textit{Entity Retrieval} and other retrieval methods in identifying retrieval documents to enhance question answering with LLMs. Figure \ref{fig:entiy_response_example} provides an intuitive example to motivate the effectiveness of \textit{Entity Retrieval}.

\section{Experiments and Analysis}
\subsection{Setup}\label{sec:setup}
We focus on Wikipedia as the knowledge base and utilize the pre-existing BM25, DPR, and ANCE retrieval indexes in Pyserini \citep{Pyserini}. These indexes, follow established practices \citep{P17-1171, 2020.emnlp-main.550} and segments the articles into non-overlapping text blocks of 100 words, resulting in 21,015,300 passages. For dense retrievers, the passages are processed with a pre-trained context encoder, generating fixed embedding vectors stored in a FAISS index \cite{FAISSLibrary}. Our experimental entity-centric questions are encoded using the question encoder, and the top $k$ relevant passages to the encoded question are retrieved from the FAISS index. For BM25 sparse retriever, the passages are stored in a Lucene index and the questions are keyword-matched to this index.

As outlined in Section \ref{sec:el_for_qa}, the document retrieval process will require loading the entire index (as well as the question encoder for dense retrieval) into memory which entails significant time and memory consumption. To address this challenge, following \citet{2023.tacl-1.75}, we treat document retrieval as a pre-processing step, caching the most relevant passages for each question before conducting the question answering experiments.

For \textit{Entity Retrieval}, similar to BM25, DPR, and ANCE, we maintain document lengths at 100 words. However, our approach diverges in sourcing documents: rather than drawing from a large index of 21 million passages, we employ the salient entities within the question and retrieve their corresponding Wikipedia articles, which we then truncate to the initial 100 words. 

We conduct our Retrieval-Augmented Question Answering experiments using LLaMA 3 model\footnote{\url{https://llama.meta.com/llama3/}.}, and in all such experiments\footnote{We run our experiments on one server containing 2 RTX A6000s with 49GB GPU memory each.}, we prevent it from generating sequences longer than 10 subwords. 

We do not use any instructional question-answer pairs in the prompts of our models\footnote{Further exploration into few-shot experimental setups involving additional (context, question, answer) in-context examples is left for future investigation.}. In the Closed-book setting, the prompt includes only the question, along with a simple instruction to answer it. In Retrieval-Augmented settings using BM25, DPR, and ANCE, the prompt incorporates pre-fetched retrieved documents from the corresponding retrieval index alongside the question and the instruction. Similarly, in the \textit{Entity Retrieval} settings, the prompt consists of the first $W$ words of the Wikipedia articles corresponding to the salient entities in the question. We follow \citet{2023.tacl-1.75} for question normalization and prompt formulation. Appendix \ref{appendix_a} provides the prompts, and example retrieved documents for each setting.

\subsection{Data}
We use the following datasets in our experiments\footnote{Please note that since \textit{Entity Retrieval} does not involve training, all mentioned dataset subsets (e.g., train, dev, or test) will be used for evaluation regardless of their names.}:

\textbf{EntityQuestions} \citep{2021.emnlp-main.496} is created by collecting 24 common relations (e.g., `author of' and `located in') and transforming fact triples (subject, relation, object) that contain these relations, into natural language questions using predefined templates. The dataset comprises 176,560 train, 22,068 dev, and 22,075 test question-answer pairs. To expedite our analytical experiments in this paper, given the extensive size of the dev and test sets, we constrain the question-answer pairs in these subsets to those featuring salient entities within the top 500K most linked Wikipedia pages, as suggested by \citet{2023.emnlp-main.686}. Thus, the dev and test subsets of EntityQuestions considered in our experiments consist of 4,710 and 4,741 questions, respectively.

\textbf{FactoidQA} \citep{FactoidQA} contains 2,203 hand crafted question-answer pairs derived from Wikipedia articles, with each pair accompanied by its corresponding Wikipedia source article included in the dataset. 

\textbf{StrategyQA} \citep{2021.tacl-1.21} is a complex boolean question answering dataset, constructed by presenting individual terms from Wikipedia to annotators. Its questions contain references to more than one Wikipedia entity, and necessitate implicit reasoning for binary (Yes/No) responses. The dataset comprises 5,111 answered questions initially intended for training question answering systems, with the system later tested on test set questions with unreleased answers. This training set is split into two subsets resulting in \texttt{train} and \texttt{train\_filtered} subsets containing 2,290 and 2,821 questions, respectively. 

\subsection{Evaluation}
We evaluate the performance of the retrieval methods using the following metrics; in each of which a document is considered \textit{relevant} if it contains a normalized form of the expected answer to the question:
\begin{itemize}
    \item nDCG@$k$ \citep[normalized Discounted Cumulative Gain at rank $k$;][]{nDCG} evaluates the quality of a ranking system by considering both the relevance and the position of documents in the top $k$ results. Mathematically, it is represented as  $$\text{nDCG@}k = \frac{\sum_{i=1}^{k} \frac{2^{r_i} - 1}{\log_2(i + 1)}}{\sum_{i=1}^{|REL_k|} \frac{2^{r_i} - 1}{\log_2(i + 1)}}$$ Where, $r_i$ denotes the relevance score of a document at the $i$\textsuperscript{th} position for a question, with relevance score $r_i = 1$ if the document is \textit{relevant}, and $r_i = 0$, otherwise. $REL_k$ refers to the \textit{relevant} subset of the retrieved documents. nDCG@$k$ scores range between 0 and 1, where a score of 1 signifies an optimal ranking with the most \textit{relevant} documents positioned at the top.
    \item MRR \citep[Mean Reciprocal Rank;][]{MRR} is the average of the reciprocal ranks of the first \textit{relevant} document for each question. Mathematically, it is represented as $$\text{MRR} = \frac{1}{|Q|} \sum_{j=1}^{|Q|} \frac{1}{r_j}$$ where $|Q|$ represents the total number of questions and $r_j$ denotes the rank of the first \textit{relevant} document for the $j$-th question.
    \item Top-$k$ Retrieval Accuracy, as reported by \citet{2021.emnlp-main.496}, is calculated as the number of questions with at least one \textit{relevant} document in the top $k$ retrieved documents divided by the total number of questions in the dataset.
\end{itemize}

We evaluate the performance of the Retrieval-Augmented Question Answering models with each retrieval method as follows:
\begin{itemize}
    \item For FactoidQA and EntityQuestions datasets, we use \texttt{OpenQA-eval} \citep{2023.acl-long.307} scripts to evaluate model performance, and report exact match (EM) and F1 scores by comparing expected answers to normalized model responses.
    \item For StrategyQA, we present accuracy scores by comparing model responses to the expected boolean answers in the dataset. As well, to assess model comprehension of the task, we count the number of answers that deviate from Yes or No and report this count in a distinct column labeled ``\texttt{Inv \#}'' for each experiment.
\end{itemize}

\begin{figure*}[ht]
    \centering
    \begin{subfigure}[b]{0.45\textwidth}
        \centering
        \scalebox{0.8}{
            \begin{tikzpicture}
                \begin{axis}[
                    xlabel={\# Retrieved Documents ($k$)},
                    ylabel={nDCG@$k$},
                    xtick={0, 1, 2, 3, 4, 5, 6, 7},
                    xticklabels={0, 1, 2, 3, 4, 5, 20, 100},
                    legend pos=south east,
                    legend columns=2,
                    grid=minor,
                    width=10cm,
                    height=7cm
                ]
                
                \addplot[color=blue, mark=*, mark size=1pt, very thick] coordinates {
                    (1, 0.1993) (2, 0.1925) (3, 0.1902) (4, 0.188) (5, 0.188) (6, 0.2015) (7, 0.2871)
                };
                \addlegendentry{BM25}
                
                \addplot[color=red, mark=*, mark size=1pt, very thick] coordinates {
                    (1, 0.1525) (2, 0.1548) (3, 0.1563) (4, 0.1576) (5, 0.1567) (6, 0.176) (7, 0.2682)
                };
                \addlegendentry{DPR}
                
                \addplot[color=green, mark=*, mark size=1pt, very thick] coordinates {
                    (1, 0.1652) (2, 0.1675) (3, 0.1694) (4, 0.1704) (5, 0.1711) (6, 0.1863) (7, 0.2777)
                };
                \addlegendentry{ANCE}
                
                \addplot[color=black, mark=star, mark size=3.5pt, thick] coordinates {(1, 0.0971)};
                \addlegendentry{ER50w}
                \draw[gray, dashed, very thick] (axis cs:0,0.0971) -- (axis cs:8,0.0971);
                
                \addplot[color=black, mark=diamond*, mark size=3pt, thick] coordinates {(1, 0.1307)};
                \addlegendentry{ER100w}
                \draw[gray, dashed, very thick] (axis cs:0,0.1307) -- (axis cs:8,0.1307);
                
                \addplot[color=black, mark=triangle*, mark size=3pt, thick] coordinates {(1, 0.1847)};
                \addlegendentry{ER300w}
                \draw[gray, dashed, very thick] (axis cs:0,0.1847) -- (axis cs:8,0.1847);
                
                \addplot[color=black, mark=square*, mark size=2.5pt, thick] coordinates {(1, 0.2719)};
                \addlegendentry{ER1000w}
                \draw[gray, dashed, very thick] (axis cs:0,0.2719) -- (axis cs:8,0.2719);
                
                \addplot[color=white] coordinates {(1, 0.072)};
                
                \end{axis}
            \end{tikzpicture}
        }
        \caption{FactoidQA}
    \end{subfigure}
    \hfill
    \begin{subfigure}[b]{0.5\textwidth}
        \centering
        \scalebox{0.8}{
            \begin{tikzpicture}
                \begin{axis}[
                    xlabel={\# Retrieved Documents ($k$)},
                    ylabel={nDCG@$k$},
                    xtick={0, 1, 2, 3, 4, 5, 6, 7},
                    xticklabels={0, 1, 2, 3, 4, 5, 20, 100},
                    legend pos=north east,
                    legend columns=2,
                    grid=minor,
                    width=10cm,
                    height=7cm
                ]
                
                \addplot[color=blue, mark=*, mark size=1pt, very thick] coordinates {
                    (1, 0.4225) (2, 0.4084) (3, 0.3961) (4, 0.386) (5, 0.3793) (6, 0.3911) (7, 0.5526)
                };
                \addlegendentry{BM25}
                
                \addplot[color=red, mark=*, mark size=1pt, very thick] coordinates {
                    (1, 0.3728) (2, 0.3643) (3, 0.3508) (4, 0.3439) (5, 0.3407) (6, 0.354) (7, 0.4768)
                };
                \addlegendentry{DPR}
                
                \addplot[color=green, mark=*, mark size=1pt, very thick] coordinates {
                    (1, 0.4512) (2, 0.4354) (3, 0.4194) (4, 0.4113) (5, 0.405) (6, 0.4168) (7, 0.5408)
                };
                \addlegendentry{ANCE}
                
                \addplot[color=black, mark=star, mark size=3.5pt, thick] coordinates {(1, 0.4346)};
                \addlegendentry{ER50w}
                \draw[gray, dashed, very thick] (axis cs:0,0.4346) -- (axis cs:8,0.4346);
                
                \addplot[color=black, mark=diamond*, mark size=3pt, thick] coordinates {(1, 0.5161)};
                \addlegendentry{ER100w}
                \draw[gray, dashed, very thick] (axis cs:0,0.5161) -- (axis cs:8,0.5161);
                
                \addplot[color=black, mark=triangle*, mark size=3pt, thick] coordinates {(1, 0.61)};
                \addlegendentry{ER300w}
                \draw[gray, dashed, very thick] (axis cs:0,0.61) -- (axis cs:8,0.61);
                
                \addplot[color=black, mark=square*, mark size=2.5pt, thick] coordinates {(1, 0.6945)};
                \addlegendentry{ER1000w}
                \draw[gray, dashed, very thick] (axis cs:0,0.6945) -- (axis cs:8,0.6945);
                
                \end{axis}
            \end{tikzpicture}
        }
        \caption{EntityQuestions (dev)}
    \end{subfigure}
    \caption{nDCG@$k$ scores evaluate the quality of BM25, DPR, ANCE, and \textit{Entity Retrieval} by considering both the relevance and the position of documents in the top $k$ retrieved passages for each question. Note that \textit{Entity Retrieval} typically results in $k$=1 document since the datasets under study often have one salient entity. The horizontal lines aid in visually comparing the performance of \textit{Entity Retrieval}, which averages one document, to other methods retrieving $k$>1 documents.}
    \label{fig:ndcg_factoidqa_and_entityquestions_dev}
    \vspace{-0.45cm}
\end{figure*}
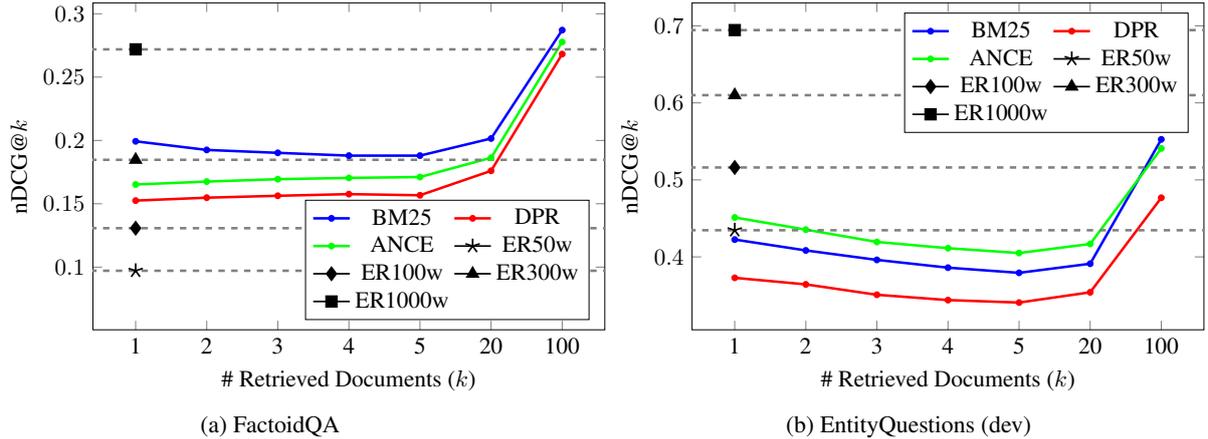

\subsection{Entity Retrieval Performance using Question Entity Annotations}
We begin our analysis by comparing \textit{Entity Retrieval} performance to BM25, DPR, and ANCE. For this experiment, we calculate nDCG with various retrieved document sets of size  $k=$ 1, 2, 3, 4, 5, 20, and 100. We use the entity annotations provided with the questions from FactoidQA and the dev set of EntityQuestions to fetch their corresponding Wikipedia articles, excluding StrategyQA from our analysis as it does not include entity annotations. On average, FactoidQA and EntityQuestions datasets contain one salient entity per question.

Apart from a few questions, the majority of FactoidQA questions, and all questions in the EntityQuestions dataset, contain only one entity annotation (leading to one augmentation document). This puts \textit{Entity Retrieval} at a disadvantage. To address this, we consider truncating the \textit{Entity Retrieval} documents to varying lengths. We compare \textit{Entity Retrieval} using the first 100 words (equivalent to the size of documents returned by BM25, DPR, and ANCE, noted as \textit{ER100w}) and also consider the first 50, 300, and 1000 words of the retrieved Wikipedia articles (noted as \textit{ER50w}, \textit{ER300w}, and \textit{ER1000w}). A 300-word \textit{Entity Retrieval} document matches the word count of three documents returned by BM25 or DPR.

Figure \ref{fig:ndcg_factoidqa_and_entityquestions_dev} presents the computed nDCG@$k$ scores across varying document sizes, highlighting the superior performance of \textit{Entity Retrieval} over other retrieval methods in the context of the entity-centric datasets under study. Notably, \textit{ER1000w}, which corresponds to ten BM25 retrieved passages in terms of word count, exhibits a retrieval performance on par with 100 retrieved documents in FactoidQA and surpasses BM25, the top-performing retriever on EntityQuestions, by 25\%. This impressive performance by \textit{Entity Retrieval} can be attributed to its ability to retrieve fewer, yet more relevant, documents. This observation aligns with the conclusion drawn by \citet{arXiv:2401.14887}, which emphasizes that the retrieval of irrelevant documents can negatively impact performance. \textit{Entity Retrieval} effectively minimizes the retrieval of such documents. Further insights can be gleaned from the comparison of nDCG scores along the x-axis of the plots in Figure \ref{fig:ndcg_factoidqa_and_entityquestions_dev}. As the number of retrieved documents increases, the likelihood of retrieving irrelevant documents also rises, leading to a decline in retrieval performance when moving from 1 to 5 retrieved documents.

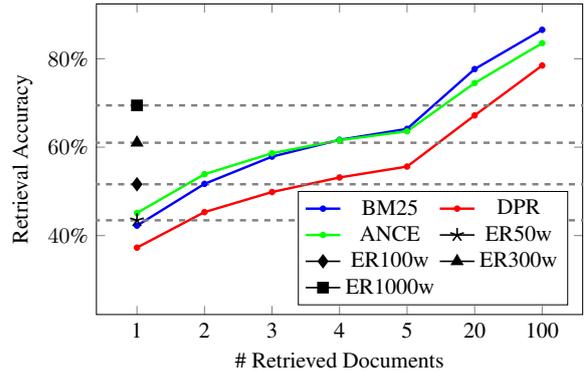
\begin{figure}
    \centering
    \centering
    \scalebox{0.76}{
        \begin{tikzpicture}
            \begin{axis}[
                xlabel={\# Retrieved Documents},
                ylabel={Retrieval Accuracy},
                xtick={0, 1, 2, 3, 4, 5, 6, 7},
                xticklabels={0, 1, 2, 3, 4, 5, 20, 100},
                yticklabel={\pgfmathparse{\tick}\pgfmathprintnumber{\pgfmathresult}\%},
                legend pos=south east,
                legend columns=2,
                grid=minor,
                width=10cm,
                height=7cm,
            ]
            
            \addplot[color=blue, mark=*, mark size=1pt, very thick] coordinates {
                (1, 42.25) (2, 51.7) (3, 57.86) (4, 61.66) (5, 64.14) (6, 77.66) (7, 86.54)
            };
            \addlegendentry{BM25}
            
            \addplot[color=red, mark=*, mark size=1pt, very thick] coordinates {
                (1, 37.28) (2, 45.31) (3, 49.85) (4, 53.15) (5, 55.61) (6, 67.18) (7, 78.47)
            };
            \addlegendentry{DPR}
            
            \addplot[color=green, mark=*, mark size=1pt, very thick] coordinates {
                (1, 45.12) (2, 53.89) (3, 58.62) (4, 61.59) (5, 63.59) (6, 74.5) (7, 83.52)
            };
            \addlegendentry{ANCE}
            
            \addplot[color=black, mark=star, mark size=3.5pt, thick] coordinates {(1, 43.46)};
            \addlegendentry{ER50w}
            \draw[gray, dashed, very thick] (axis cs:0,43.46) -- (axis cs:8,43.46);
            
            \addplot[color=black, mark=diamond*, mark size=3pt, thick] coordinates {(1, 51.61)};
            \addlegendentry{ER100w}
            \draw[gray, dashed, very thick] (axis cs:0,51.61) -- (axis cs:8,51.61);
            
            \addplot[color=black, mark=triangle*, mark size=3pt, thick] coordinates {(1, 61)};
            \addlegendentry{ER300w}
            \draw[gray, dashed, very thick] (axis cs:0,61) -- (axis cs:8,61);
            
            \addplot[color=black, mark=square*, mark size=2.5pt, thick] coordinates {(1, 69.45)};
            \addlegendentry{ER1000w}
            \draw[gray, dashed, very thick] (axis cs:0,69.45) -- (axis cs:8,69.45);
            
            \addplot[color=white] coordinates {(1, 28)};
            
            \end{axis}
            \end{tikzpicture}
    }
    \caption{Retrieval Accuracy scores showcasing the correlation between the number of retrieved documents and the expected answers' coverage in EntityQuestions (dev) subset.}
    \label{fig:racc_entityquestions_dev}
    \vspace{-0.4cm}
\end{figure}

Table \ref{tab:mrr_factoidqa_and_entityquestions_dev} showcases the calculated MRR scores, emphasizing the quicker attainment of relevant retrieval documents in \textit{Entity Retrieval} compared to other retrieval methods. Concurrently, Figure \ref{fig:racc_entityquestions_dev} illustrates the impact of incrementing the number of retrieved documents on the expansion of the expected answers' coverage for the EntityQuestions dev subset.

\begin{table}[ht]
    \centering
    \setlength{\tabcolsep}{2pt}
    \begin{tabular}{lcc}
        \toprule
        ~ & \textbf{FactoidQA} & \textbf{EntityQuestions (dev)} \\
        \midrule
        BM25 & 0.245 & 0.522 \\
        DPR & 0.209 & 0.456 \\
        ANCE & 0.222 & 0.536 \\\midrule
        ER50w & 0.097 & 0.435 \\
        ER100w & 0.131 & 0.516 \\
        ER300w & 0.185 & 0.610 \\
        ER1000w & \textbf{0.272} & \textbf{0.695} \\
        \bottomrule
    \end{tabular}
    \caption{MRR scores comparing the retrieval quality of BM25, DPR, ANCE, and \textit{Entity Retrieval} through the average of the reciprocal ranks of the first relevant document for each question.}
    \label{tab:mrr_factoidqa_and_entityquestions_dev}
    \vspace{-0.4cm}
\end{table}

While it may be appealing to consider 100 or more documents to simultaneously enhance both nDCG and Retrieval Accuracy, it is important to note that 100 retrieved documents would comprise 10,000 words. This could potentially overwhelm the model with excessive noise (irrelevant documents), and as well, could make it extremely costly to execute Retrieval-Augmented Question Answering, especially when the cost of API calls is calculated per token. We would need at least 10,000 tokens (optimistically, assuming each word equates to only one token) in addition to the tokens in the question. These factors suggest that retrieving a few documents for each question is more beneficial.

Taking these considerations into account, along with the nDCG@$k$, MRR, and Retrieval Accuracy results from this section, we gain a comprehensive understanding of the trade-off between the quality of the retrieved documents, which diminishes as we consider more documents, and the answer coverage, which increases as the model has a higher chance of encountering the right document with the correct hint for the answer. Consequently, we opt for $k=4$ as a default, and we will always retrieve the top-4 documents in our Retrieval-Augmented Question Answering experiments.

\subsection{Retrieval-Augmented Question Answering}\label{sec:raqa_results}

Next, we examine the effectiveness of our proposed \textit{Entity Retrieval} method compared to other retrieval methods in improving the quality of responses to entity-centric questions. We explore three settings: Closed-book, Retrieval-Augmented, and \textit{Entity Retrieval} with question entity annotations (Section \ref{sec:setup}). The primary purpose of using question entity annotations is to demonstrate their ability to accurately identify relevant augmentation documents. These experiments establish an expected performance ceiling for \textit{Entity Retrieval} and can inspire future research to meet or exceed this threshold.

\begin{table*}[ht]
	\centering
	\begin{tabular}{l|cc|cc|cc}
	\toprule
	\multicolumn{1}{c|}{\multirow{3}{*}{\begin{tabular}[c]{@{}c@{}}\textbf{LLaMA3} \\ \textbf{(8B)}\end{tabular}}} & \multicolumn{2}{c|}{\multirow{2}{*}{\textbf{FactoidQA}}} & \multicolumn{4}{c}{\textbf{EntityQuestions}}                    \\ \cmidrule{4-7} 
	\multicolumn{1}{c|}{}                             & \multicolumn{2}{c|}{}                           & \multicolumn{2}{c|}{\textbf{dev}}     & \multicolumn{2}{c}{\textbf{test}} \\ \cmidrule{2-7} 
	\multicolumn{1}{c|}{}                             & \textbf{EM}           & \multicolumn{1}{c|}{\textbf{F1}}          & \textbf{EM} & \multicolumn{1}{c|}{\textbf{F1}} & \textbf{EM}          & \textbf{F1}         \\ \midrule

	\multicolumn{1}{l|}{Closed-book}  & 30.5\small{$\pm$0.4} & 39.3\small{$\pm$0.0}  & 22.9\small{$\pm$0.5} & 37.9\small{$\pm$0.7}  & 22.9\small{$\pm$0.2} & 38.3\small{$\pm$0.5} \\\midrule
	\multicolumn{7}{c}{Retrieval-Augmented QA} \\ \midrule
	\multicolumn{1}{l|}{BM25}         & 32.4\small{$\pm$0.8} & 42.6\small{$\pm$0.3}  & 23.7\small{$\pm$0.3} & 38.5\small{$\pm$0.6}  & 23.4\small{$\pm$0.2} & 38.7\small{$\pm$0.3} \\
	\multicolumn{1}{l|}{DPR}          & 29.8\small{$\pm$1.0} & 38.9\small{$\pm$1.1}  & 21.9\small{$\pm$0.3} & 36.2\small{$\pm$0.2}  & 20.7\small{$\pm$0.6} & 35.4\small{$\pm$0.4} \\
	\multicolumn{1}{l|}{ANCE}         & 30.4\small{$\pm$0.4} & 39.9\small{$\pm$0.3}  & 23.1\small{$\pm$0.5} & 37.9\small{$\pm$0.4}  & 22.7\small{$\pm$0.5} & 37.9\small{$\pm$0.6} \\\midrule
	\multicolumn{7}{c}{\textit{Entity Retrieval} w/ Question Entity Annotations} \\ \midrule
	\multicolumn{1}{l|}{ER50w}        & 34.4\small{$\pm$0.5} & 43.7\small{$\pm$0.5}  & 24.9\small{$\pm$0.1} & 41.2\small{$\pm$0.1}  & 24.1\small{$\pm$0.6} & 41.1\small{$\pm$0.3} \\
	\multicolumn{1}{l|}{ER100w}       & 33.6\small{$\pm$0.3} & 42.9\small{$\pm$0.4}  & \textbf{26.3}\small{$\pm$0.2} & \textbf{42.8}\small{$\pm$0.1}  & \textbf{25.7}\small{$\pm$0.1} & \textbf{42.4}\small{$\pm$0.0} \\
	\multicolumn{1}{l|}{ER300w}       & 33.7\small{$\pm$0.9} & 43.0\small{$\pm$1.1}  & 26.2\small{$\pm$0.3} & 42.7\small{$\pm$0.1}  & 25.5\small{$\pm$0.7} & \textbf{42.4}\small{$\pm$0.8} \\
	\multicolumn{1}{l|}{ER1000w}      & \textbf{35.0}\small{$\pm$0.3} & \textbf{44.9}\small{$\pm$0.5}  & 25.1\small{$\pm$0.4} & 41.9\small{$\pm$0.4}  & 24.2\small{$\pm$0.9} & 41.1\small{$\pm$0.6} \\\midrule
	\multicolumn{7}{c}{\textit{Entity Retrieval} w/ \textsc{SpEL} Entity Annotations} \\ \midrule
	\multicolumn{1}{l|}{ERSp50w}      & 29.6\small{$\pm$0.3} & 38.6\small{$\pm$0.5}  & 24.1\small{$\pm$0.5} & 39.1\small{$\pm$0.2}  & 23.6\small{$\pm$0.8} & 39.4\small{$\pm$0.5} \\
	\multicolumn{1}{l|}{ERSp100w}     & 28.7\small{$\pm$0.9} & 37.7\small{$\pm$1.0}  & 24.8\small{$\pm$0.5} & 40.0\small{$\pm$0.2}  & 24.4\small{$\pm$0.3} & 39.9\small{$\pm$0.2} \\
	\multicolumn{1}{l|}{ERSp300w}     & 26.9\small{$\pm$0.4} & 35.6\small{$\pm$0.5}  & 24.5\small{$\pm$0.3} & 39.9\small{$\pm$0.4}  & 24.4\small{$\pm$0.5} & 40.2\small{$\pm$0.3} \\
	\multicolumn{1}{l|}{ERSp1000w}    & 21.7\small{$\pm$0.7} & 30.8\small{$\pm$1.0}  & 24.2\small{$\pm$0.2} & 39.6\small{$\pm$0.3}  & 22.9\small{$\pm$0.5} & 39.0\small{$\pm$0.7} \\
	\bottomrule
	\end{tabular}
	\caption{Question answering efficacy comparison between Closed-book and Retrieval-Augmentation using BM25, DPR, ANCE, and \textit{Entity Retrieval}. EM refers to the exact match between predicted and expected answers, disregarding punctuation and articles (\texttt{a}, \texttt{an}, \texttt{the}).\\\textsuperscript{$\star$} Results represent the average of three runs, accompanied by a margin of error based on a 99\% confidence interval.}
	\label{tab:llama_3_8b_raqa_results}
 \vspace{-0.2cm}
\end{table*}

The initial eight rows of Table \ref{tab:llama_3_8b_raqa_results} present the results of our experiments using LLaMA 3 (8B) model. Upon examining these results, it is evident that \textit{ER100w}, the most analogous \textit{Entity Retrieval} setting to other retrieval methods, outperforms in terms of both EM and F1 scores. This setting, like the other retrieval methods, returns 100-word documents. However, as we noted earlier, \textit{Entity Retrieval} generally retrieves fewer documents overall, making it both more accurate and more efficient. 

Our dense retrieval results align with the observations of \citet{2021.emnlp-main.496}, asserting that entity-centric questions indeed challenge dense retrievers. Although the BM25 method proves successful in enhancing the results compared to the Closed-book setting, it is noteworthy that even \textit{Entity Retrieval} with the initial 50 words of the articles corresponding to the salient entities within questions yields superior results. This is particularly significant when compared to other retrieval methods which necessitate indexing the entire knowledge base on disk and loading the index into memory; a process required in inference time where caching is not an option.

\subsection{Entity Retrieval in absence of Question Entity Annotations}

Section \ref{sec:raqa_results} establishes \textit{Entity Retrieval} as a viable augmentation method for entity-centric questions. Next, we aim to reach the established performance ceiling in the absence of question entity annotations. Here, we examine the potential of entity linking as an automated method to provide these annotations. Our primary research question is: how effectively can current entity linking methods help \textit{Entity Retrieval} achieve optimal performance?

Ideally, we would like to evaluate all recent entity linking methods to identify the most effective one. However, due to time and budget limitations, we depend on the recent benchmarking studies by \citet{2024.naacl-main.???} to choose a method. They examine the latest entity linking methods in terms of performance against unseen data and endorse \textsc{SpEL} \citep{2023.emnlp-main.686} as the top performer. Consequently, we investigate \textit{Entity Retrieval} using entities identified with \textsc{SpEL}, while reserving the examination of other entity linking techniques for \textit{Entity Retrieval} for future research. 

We maintain the \textit{Entity Retrieval} settings as before, defining \textit{ERSp50w}, \textit{ERSp100w}, \textit{ERSp300w}, and \textit{ERSp1000w} for performing entity linking with \textsc{SpEL}, then retrieving the Wikipedia articles corresponding to the \textsc{SpEL} identified entities, and using the first 50, 100, 300, and 1000 words of these articles as documents to augment the question when prompting the LLM. Table \ref{tab:spel_annotation_stats} presents the aggregated entity identification statistics of \textsc{SpEL} across various subsets of each dataset under study.

\begin{table}[t]
    \centering
    \begin{tabular}{l|ccc}
        \toprule
                         & Max. & Avg.  & Linked \%\\\midrule 
         FactoidQA       & 8    &  0.8  & 56.5\% \\
         EntityQuestions & 3    &  0.7  & 65.6\% \\ 
         StrategyQA      & 4    &  1.1  & 74.9\%\\
         \bottomrule
    \end{tabular}
    \caption{Maximum and Average \textsc{SpEL} identified entity count as well as the total percentage of questions with at least one identified entity in each dataset. \textsc{SpEL} successfully identifies and links entities in 1,244 FactoidQA, 3,108 EntityQuestions (dev), 3,095 EntityQuestions (test), 1,735 StrategyQA (train), and 2,094 StrategyQA (train\_filtered) questions. For the remaining questions in each dataset where no entities are identified, they will be introduced to the LLM without any augmented documents in the \textit{Entity Retrieval} settings.}
    \label{tab:spel_annotation_stats}
    \vspace{-0.4cm}
\end{table}

\begin{table*}
    \centering
    \scalebox{0.85}{
    \begin{tabular}{l|l|l}
        \toprule
        Question    & Who performed Alexis Colby?                 & What is the capital of Seine-Saint-Denis? \\
        Answer      & Joan Collins                                & Bobigny \\\midrule
        Closed-Book & Diana Ross                                  & Paris \\
        BM25        & Linda Evans                                 & Saint-Denis \\
        DPR         & Alexis Cohen                                & Saint-Denis \\
        ANCE        & Nicollette Sheridan performed Alexis Colby. & Saint-Denis \\
        ERSp100w    & Joan Collins                                & Bobigny \\
        \midrule\midrule
        Question    & Where did John Snetzler die?                & Where was Brigita Bukovec born? \\
        Answer      & Schaffhausen                                & Ljubljana \\\midrule
        Closed-Book & He died in London, England, in 178          & Brigita Bukovec was born in Slovenia \\
        BM25        & John Snetzler died in London.               & Slovenia \\
        DPR         & John Snetzler died in London                & in Slovakia \\
        ANCE        & in England                                  & Rîbnița \\
        ERSp100w    & Schaffhausen                                & Ljubljana \\
        \bottomrule
    \end{tabular}
    }
    \caption{Example questions from EntityQuestions (dev) to demonstrate the performance of \textit{Entity Retrieval}.}
    \label{tab:raqa_comparison_examples}
    \vspace{-0.35cm}
\end{table*}

\begin{table*}[ht]
    \centering
    \begin{tabular}{l|ll|ll}
	\toprule
	\multicolumn{1}{c|}{\multirow{3}{*}{\begin{tabular}[c]{@{}c@{}}\textbf{LLaMA3} \\ \textbf{(8B)}\end{tabular}}} & \multicolumn{2}{c|}{\textbf{train}}                            & \multicolumn{2}{c}{\textbf{train\_filtered}}                  \\ \cmidrule{2-5} 
	\multicolumn{1}{c|}{} & \multicolumn{1}{c}{\textbf{Acc.}} & \multicolumn{1}{c|}{\textbf{Inv \#}} & \multicolumn{1}{c}{\textbf{Acc.}} & \multicolumn{1}{c}{\textbf{Inv \#}} \\ \midrule
	\multicolumn{1}{l|}{BM25}      & 43.5\small{$\pm$0.6} & 608\small{$\pm$14}  & 48.9\small{$\pm$0.7} & 673\small{$\pm$12}  \\
	\multicolumn{1}{l|}{ANCE}      & 46.6\small{$\pm$1.3} & 552\small{$\pm$11}  & 51.8\small{$\pm$0.7} & 647\small{$\pm$35}  \\\midrule
	\multicolumn{1}{l|}{ERSp50w}   & 50.1\small{$\pm$1.1} & 370\small{$\pm$28}  & \textbf{56.3}\small{$\pm$0.9} & 417\small{$\pm$21}  \\
	\multicolumn{1}{l|}{ERSp100w}  & \textbf{50.3}\small{$\pm$1.4} & \textbf{369}\small{$\pm$15}  & {56.2}\small{$\pm$0.8} & \textbf{384}\small{$\pm$9}  \\
	\multicolumn{1}{l|}{ERSp300w}  & 46.2\small{$\pm$1.3} & 504\small{$\pm$17} & 53.5\small{$\pm$1.5} & 546\small{$\pm$20}  \\
	\multicolumn{1}{l|}{ERSp1000w} & 39.5\small{$\pm$1.4} & 775\small{$\pm$6}  & 43.4\small{$\pm$0.5} & 919\small{$\pm$14}  \\
	\bottomrule
	\end{tabular}
    \caption{Comparison of \textit{Entity Retrieval} using \textsc{SpEL} identified entities to the best-performing dense and sparse retrieval methods of Table \ref{tab:llama_3_8b_raqa_results} on the StrategyQA dataset. Given the expected boolean results for StrategyQA questions, we restricted LLaMA 3 to generate only one token. \textit{Acc.} indicates the fraction of answers that correctly match the expected Yes or No responses in the dataset, while \textit{Inv \#} represents the count of labels that are neither Yes nor No, but another invalid answer.\\\textsuperscript{$\star$} Results represent the average of three runs, accompanied by a margin of error based on a 99\% confidence interval.}
    \label{tab:llama_3_8b_strategyqa_results}
    \vspace{-0.35cm}
\end{table*}

The final four rows of Table \ref{tab:llama_3_8b_raqa_results} showcase the comparative results of utilizing entities identified by \textsc{SpEL} for \textit{Entity Retrieval}. Given that one-third of EntityQuestions and approximately half of FactoidQA lack identified annotations, the exact match scores reveal that \textit{Entity Retrieval} performs robustly and surpasses BM25, the top-performing competitor, for EntityQuestions while approaching DPR's performance for FactoidQA. This underscores the potential of \textit{Entity Retrieval} within this paradigm. In addition, the disparity between the results with and without question entity annotations strongly indicates the necessity for further research in Entity Linking, which could enhance entity-centric question answering as a downstream task. Table \ref{tab:raqa_comparison_examples} provides some example questions where \textit{Entity Retrieval} has led to better answers.

Table \ref{tab:llama_3_8b_strategyqa_results} compares of the performance of \textit{Entity Retrieval} using \textsc{SpEL} identified entities against other retrieval methods on the StrategyQA dataset. The results clearly demonstrate the superior performance of \textit{Entity Retrieval} over the top-performing retrieval methods of Table \ref{tab:llama_3_8b_raqa_results}. It is important to note that the 100-word setting (\textit{ERSp100w}) is the most analogous to other retrieval methods. Interestingly, the results from the 1000-word setting suggest that longer documents do not necessarily enhance the model's recall. In fact, beyond a certain length, the model may become overwhelmed by the sheer volume of noise, leading to confusion. Lastly, the invalid count values suggest that \textit{Entity Retrieval} is more effective in assisting the model to comprehend the boolean nature of expected responses, eliminating the need to rely on retrieval from millions of passages.

\subsection{Real-time Efficiency Analysis}\label{sec:raqa_efficiency_analysis}

Our analysis thus far has primarily focused on the retrieval performance, without consideration for the time and memory efficiency; crucial factors in retrieval method selection. In this section, we shift our focus to these aspects. 

We begin by replacing our pre-built retrieval cache document sets with the original retrieval modules that were used in creating the cached sets. We load the indexes and the necessary models for fetching the retrieval documents. We then record the peak main memory requirement of each method during the experiment. It is important to note that all retrieval methods primarily rely on main memory, with minimal differences in GPU memory requirements. Therefore, we report an average GPU memory requirement of 35GB for LLaMA 3 (8B) and exclude it from our results table. We then feed all 2,203 FactoidQA questions into the BM25, ANCE, and \textit{Entity Retrieval} (using \textsc{SpEL} identified entities) to fetch the top-4 documents. We report the total time taken to generate answers to all the questions, which includes the time for querying the BM25 or ANCE indexes in the Retrieval-Augmented settings, or the time for performing on-the-fly entity linking and fetching the Wikipedia articles from disk in the \textit{Entity Retrieval} setting. Additionally, we keep track of all the pre-built models and indexes that each method requires for download and storage. We report the total size of all downloaded files to disk.

\begin{table}
\centering
\begin{tabular}{lrrr}
\toprule
\multicolumn{1}{c}{\textbf{}} & \multicolumn{1}{c}{\textbf{\begin{tabular}[c]{@{}c@{}}Total\\Time\end{tabular}}} & \multicolumn{1}{c}{\textbf{\begin{tabular}[c]{@{}c@{}}Disk\\Storage\end{tabular}}} & \multicolumn{1}{c}{\textbf{\begin{tabular}[c]{@{}c@{}}Main\\Memory\end{tabular}}} \\
\midrule
BM25      &  45min &   11GB & 2.3GB \\ 
ANCE      & 960min & 61.5GB & 64.2GB\\ 
ERSp100w  &  34min &  9.4GB & 6.3GB \\
\bottomrule
\end{tabular}
\caption{Comparison of the required resources for each retrieval method in real-time execution. The reported total time values exclude the time taken to load the indexes and models, focusing solely on the time used to answer the questions.}
\label{tab:efficieny_analysis}
\vspace{-0.4cm}
\end{table}

Table \ref{tab:efficieny_analysis} presents our findings on time and memory requirements. It is evident that ANCE requires significantly more time to fetch and provide documents, six times more disk space to store its indexes, and over ten times higher main memory demands to load its dense representations\footnote{Our empirical results demonstrate that DPR follows the same trend.}. In contrast, BM25 and \textit{Entity Retrieval} are more resource-friendly. Notably, \textit{Entity Retrieval} is 25\% faster than BM25 in response generation while demanding the total memory and disk space of a standard personal computer. Future research can be directed towards reducing the memory requirements of \textit{Entity Retrieval}; a direction which we find quit promising.

\section{Related Work}
Similar to our studies, \citet{LLMLearnLongTail?} investigate the impact of salient entities on question answering, and propose constructing oracle retrieval documents as the 300-word segment surrounding the ground-truth answer from the Wikipedia page that contains the answer (entity name). Our approach leverages salient entities from questions without directly involving answers. Additionally, they primarily use entities to classify questions into those concerning frequent knowledge base entries versus those about rare entries on the long-tail, whereas our approach assigns a more substantial role to entities, treating them as pointers guiding the retrieval of relevant documents to augment questions.

\citet{2021.emnlp-main.496} compare DPR and BM25 retrievers for entity-centric questions, and demonstrate that DPR greatly underperforms BM25. They attribute this to dense retrievers' difficulty with infrequent entities, which are less represented in training data. In contrast, BM25's frequency-based retrieval is not sensitive to entity frequency. We take a parallel approach and propose a simple yet effective method that leverages salient entities in the question for identifying augmentation documents. 

Similar to our studies, \citet{DifferentiableReasoning, Learning2Retrieve} focus on answering questions with minimal lexical overlap between the retrieved documents and the question text. However, they emphasize multi-hop question answering, using entity linking to extract entities from the question and leveraging knowledge base articles to guide the multi-hop process. In contrast, we utilize entity links to directly identify augmentation documents. \citet{D18-1455} employ entity linking to identify entities in the question, generating a set of seed entities, which are then expanded using the PPR algorithm to create a subgraph of the knowledge base containing relevant entities. A graph propagation algorithm subsequently learns representations for each node in the subgraph, and each representation is binary classified to determine if it answers the question. Our approach differs as we focus on using LLMs, employing entity linking in a Retrieval-Augmented setting without relying on graph propagation.

\section{Conclusion}

In this study, we focused on Retrieval-Augmented Question Answering, and explored various retrieval methods that rely on the similarity between the question and the content of the passages to be retrieved. We introduced a novel approach, \textit{Entity Retrieval}, which deviates from the conventional textual similarity-based mechanism. Instead, it capitalizes on the salient entities within the question to identify retrieval documents. Our findings indicate that our proposed method is not only more accurate but also faster in the context of entity-centric question answering. 


\section*{Limitations and Ethical Considerations}

Our proposed \textit{Entity Retrieval} method is specifically tailored for answering entity-centric questions, with its performance heavily reliant on the presence of question entities. In scenarios where entity annotations are absent, the method's effectiveness is directly tied to the performance of external entity linking methods. We acknowledge that our exploration of potential entity linking methods has not been exhaustive, and further investigation may yield insights that could enhance the \textit{Entity Retrieval} method, even in the absence of question entity annotations.

Furthermore, we recognize that entity linking can occasionally result in ambiguous entities. Our research has not delved into the impact of such ambiguities on the \textit{Entity Retrieval} method, and we propose that future studies should focus on ensuring the selection of the most contextually appropriate entities for retrieval.

Our research is primarily centered on Wikipedia as the knowledge base, a choice heavily influenced by previous studies for the sake of comparability. However, we acknowledge the importance of exploring other knowledge bases and ontologies, particularly in different domains, such as UMLS \citep{UMLS} in the medical field.

In terms of benchmarking, we have compared the \textit{Entity Retrieval} method against a limited selection of existing retrieval methods, guided by our judgement, experience, and considerations of implementation availability. We concede that our comparison has not been exhaustive, and this reasoning extends to our comparison using different LLMs and their available sizes.

Our research is on English only, and we acknowledge that entity-centric question answering in other languages is also relevant and important. We hope to extend our work to cover multiple languages in the future. We inherit the biases that exist in the data used in this project, and we do not explicitly de-bias the data. We are providing our code to the research community and we trust that those who use the model will do so ethically and responsibly.

\bibliography{references}

\begin{thebibliography}{58}
\expandafter\ifx\csname natexlab\endcsname\relax\def\natexlab#1{#1}\fi

\bibitem[{Abney et~al.(2000)Abney, Collins, and Singhal}]{A00-1041}
Steven Abney, Michael Collins, and Amit Singhal. 2000.
\newblock \href {https://doi.org/10.3115/974147.974188} {Answer extraction}.
\newblock In \emph{Sixth Applied Natural Language Processing Conference}, pages
  296--301, Seattle, Washington, USA. Association for Computational
  Linguistics.

\bibitem[{Aghaebrahimian and Jur{\v{c}}{\'\i}{\v{c}}ek(2016)}]{W16-0104}
Ahmad Aghaebrahimian and Filip Jur{\v{c}}{\'\i}{\v{c}}ek. 2016.
\newblock \href {https://doi.org/10.18653/v1/W16-0104} {Open-domain factoid
  question answering via knowledge graph search}.
\newblock In \emph{Proceedings of the Workshop on Human-Computer Question
  Answering}, pages 22--28, San Diego, California. Association for
  Computational Linguistics.

\bibitem[{Alizadeh et~al.(2023)Alizadeh, Mirzadeh, Belenko, Khatamifard, Cho,
  Del~Mundo, Rastegari, and Farajtabar}]{arXiv:2312.11514}
Keivan Alizadeh, Iman Mirzadeh, Dmitry Belenko, Karen Khatamifard, Minsik Cho,
  Carlo~C Del~Mundo, Mohammad Rastegari, and Mehrdad Farajtabar. 2023.
\newblock \href {https://arxiv.org/pdf/2312.11514.pdf} {Llm in a flash:
  Efficient large language model inference with limited memory}.
\newblock \emph{arXiv preprint arXiv:2312.11514}.

\bibitem[{Asai et~al.(2020)Asai, Hashimoto, Hajishirzi, Socher, and
  Xiong}]{Learning2Retrieve}
Akari Asai, Kazuma Hashimoto, Hannaneh Hajishirzi, Richard Socher, and Caiming
  Xiong. 2020.
\newblock \href {https://openreview.net/forum?id=SJgVHkrYDH} {Learning to
  retrieve reasoning paths over wikipedia graph for question answering}.
\newblock In \emph{International Conference on Learning Representations}.

\bibitem[{Bodenreider(2004)}]{UMLS}
Olivier Bodenreider. 2004.
\newblock \href {https://pubmed.ncbi.nlm.nih.gov/14681409/} {The unified
  medical language system (umls): integrating biomedical terminology}.
\newblock \emph{Nucleic acids research}, 32(suppl\_1):D267--D270.

\bibitem[{Chen et~al.(2017)Chen, Fisch, Weston, and Bordes}]{P17-1171}
Danqi Chen, Adam Fisch, Jason Weston, and Antoine Bordes. 2017.
\newblock \href {https://doi.org/10.18653/v1/P17-1171} {Reading {W}ikipedia to
  answer open-domain questions}.
\newblock In \emph{Proceedings of the 55th Annual Meeting of the Association
  for Computational Linguistics (Volume 1: Long Papers)}, pages 1870--1879,
  Vancouver, Canada. Association for Computational Linguistics.

\bibitem[{Chen et~al.(2022)Chen, Lakhotia, Oguz, Gupta, Lewis, Peshterliev,
  Mehdad, Gupta, and Yih}]{2022.findings-emnlp.19}
Xilun Chen, Kushal Lakhotia, Barlas Oguz, Anchit Gupta, Patrick Lewis, Stan
  Peshterliev, Yashar Mehdad, Sonal Gupta, and Wen-tau Yih. 2022.
\newblock \href {https://doi.org/10.18653/v1/2022.findings-emnlp.19} {Salient
  phrase aware dense retrieval: Can a dense retriever imitate a sparse one?}
\newblock In \emph{Findings of the Association for Computational Linguistics:
  EMNLP 2022}, pages 250--262, Abu Dhabi, United Arab Emirates. Association for
  Computational Linguistics.

\bibitem[{Choi et~al.(2018)Choi, He, Iyyer, Yatskar, Yih, Choi, Liang, and
  Zettlemoyer}]{D18-1241}
Eunsol Choi, He~He, Mohit Iyyer, Mark Yatskar, Wen-tau Yih, Yejin Choi, Percy
  Liang, and Luke Zettlemoyer. 2018.
\newblock \href {https://doi.org/10.18653/v1/D18-1241} {{Q}u{AC}: Question
  answering in context}.
\newblock In \emph{Proceedings of the 2018 Conference on Empirical Methods in
  Natural Language Processing}, pages 2174--2184, Brussels, Belgium.
  Association for Computational Linguistics.

\bibitem[{Cuconasu et~al.(2024)Cuconasu, Trappolini, Siciliano, Filice,
  Campagnano, Maarek, Tonellotto, and Silvestri}]{arXiv:2401.14887}
Florin Cuconasu, Giovanni Trappolini, Federico Siciliano, Simone Filice, Cesare
  Campagnano, Yoelle Maarek, Nicola Tonellotto, and Fabrizio Silvestri. 2024.
\newblock \href {https://arxiv.org/pdf/2401.14887} {The power of noise:
  Redefining retrieval for rag systems}.
\newblock \emph{arXiv preprint arXiv:2401.14887}.

\bibitem[{Cui et~al.(2017)Cui, Xiao, Wang, Song, Hwang, and Wang}]{KBQA}
Wanyun Cui, Yanghua Xiao, Haixun Wang, Yangqiu Song, Seung-won Hwang, and Wei
  Wang. 2017.
\newblock \href {https://www.vldb.org/pvldb/vol10/p565-cui.pdf} {Kbqa: Learning
  question answering over qa corpora and knowledge bases}.
\newblock \emph{Proceedings of the VLDB Endowment}, 10(5).

\bibitem[{Devlin et~al.(2019)Devlin, Chang, Lee, and Toutanova}]{N19-1423}
Jacob Devlin, Ming-Wei Chang, Kenton Lee, and Kristina Toutanova. 2019.
\newblock \href {https://doi.org/10.18653/v1/N19-1423} {{BERT}: Pre-training of
  deep bidirectional transformers for language understanding}.
\newblock In \emph{Proceedings of the 2019 Conference of the North {A}merican
  Chapter of the Association for Computational Linguistics: Human Language
  Technologies, Volume 1 (Long and Short Papers)}, pages 4171--4186,
  Minneapolis, Minnesota. Association for Computational Linguistics.

\bibitem[{Dhingra et~al.(2020)Dhingra, Zaheer, Balachandran, Neubig,
  Salakhutdinov, and Cohen}]{DifferentiableReasoning}
Bhuwan Dhingra, Manzil Zaheer, Vidhisha Balachandran, Graham Neubig, Ruslan
  Salakhutdinov, and William~W. Cohen. 2020.
\newblock \href {https://openreview.net/forum?id=SJxstlHFPH} {Differentiable
  reasoning over a virtual knowledge base}.
\newblock In \emph{International Conference on Learning Representations}.

\bibitem[{Douze et~al.(2024)Douze, Guzhva, Deng, Johnson, Szilvasy, Mazaré,
  Lomeli, Hosseini, and Jégou}]{FAISSLibrary}
Matthijs Douze, Alexandr Guzhva, Chengqi Deng, Jeff Johnson, Gergely Szilvasy,
  Pierre-Emmanuel Mazaré, Maria Lomeli, Lucas Hosseini, and Hervé Jégou.
  2024.
\newblock \href {https://arxiv.org/pdf/2401.08281.pdf} {The faiss library}.
\newblock \emph{arXiv preprint arXiv:2401.08281}.

\bibitem[{Geva et~al.(2021)Geva, Khashabi, Segal, Khot, Roth, and
  Berant}]{2021.tacl-1.21}
Mor Geva, Daniel Khashabi, Elad Segal, Tushar Khot, Dan Roth, and Jonathan
  Berant. 2021.
\newblock \href {https://doi.org/10.1162/tacl_a_00370} {Did aristotle use a
  laptop? a question answering benchmark with implicit reasoning strategies}.
\newblock \emph{Transactions of the Association for Computational Linguistics},
  9:346--361.

\bibitem[{Glass et~al.(2022)Glass, Rossiello, Chowdhury, Naik, Cai, and
  Gliozzo}]{2022.naacl-main.194}
Michael Glass, Gaetano Rossiello, Md~Faisal~Mahbub Chowdhury, Ankita Naik,
  Pengshan Cai, and Alfio Gliozzo. 2022.
\newblock \href {https://doi.org/10.18653/v1/2022.naacl-main.194} {{R}e2{G}:
  Retrieve, rerank, generate}.
\newblock In \emph{Proceedings of the 2022 Conference of the North American
  Chapter of the Association for Computational Linguistics: Human Language
  Technologies}, pages 2701--2715, Seattle, United States. Association for
  Computational Linguistics.

\bibitem[{Hochreiter and Schmidhuber(1997)}]{LSTM}
Sepp Hochreiter and J{\"u}rgen Schmidhuber. 1997.
\newblock Long short-term memory.
\newblock \emph{Neural computation}, 9(8):1735--1780.

\bibitem[{Izacard et~al.(2022)Izacard, Caron, Hosseini, Riedel, Bojanowski,
  Joulin, and Grave}]{Contriever}
Gautier Izacard, Mathilde Caron, Lucas Hosseini, Sebastian Riedel, Piotr
  Bojanowski, Armand Joulin, and Edouard Grave. 2022.
\newblock \href {https://openreview.net/forum?id=jKN1pXi7b0} {Unsupervised
  dense information retrieval with contrastive learning}.
\newblock \emph{Transactions on Machine Learning Research}.

\bibitem[{Izacard and Grave(2021{\natexlab{a}})}]{DKRR}
Gautier Izacard and Edouard Grave. 2021{\natexlab{a}}.
\newblock \href {https://openreview.net/forum?id=NTEz-6wysdb} {Distilling
  knowledge from reader to retriever for question answering}.
\newblock In \emph{International Conference on Learning Representations}.

\bibitem[{Izacard and Grave(2021{\natexlab{b}})}]{2021.eacl-main.74}
Gautier Izacard and Edouard Grave. 2021{\natexlab{b}}.
\newblock \href {https://doi.org/10.18653/v1/2021.eacl-main.74} {Leveraging
  passage retrieval with generative models for open domain question answering}.
\newblock In \emph{Proceedings of the 16th Conference of the European Chapter
  of the Association for Computational Linguistics: Main Volume}, pages
  874--880, Online. Association for Computational Linguistics.

\bibitem[{J{\"a}rvelin and Kek{\"a}l{\"a}inen(2002)}]{nDCG}
Kalervo J{\"a}rvelin and Jaana Kek{\"a}l{\"a}inen. 2002.
\newblock \href {https://faculty.cc.gatech.edu/~zha/CS8803WST/dcg.pdf}
  {Cumulated gain-based evaluation of ir techniques}.
\newblock \emph{ACM Transactions on Information Systems (TOIS)},
  20(4):422--446.

\bibitem[{Johnson et~al.(2019)Johnson, Douze, and J{\'e}gou}]{FAISS}
Jeff Johnson, Matthijs Douze, and Herv{\'e} J{\'e}gou. 2019.
\newblock \href {https://arxiv.org/pdf/1702.08734.pdf} {Billion-scale
  similarity search with gpus}.
\newblock \emph{IEEE Transactions on Big Data}, 7(3):535--547.

\bibitem[{Kamalloo et~al.(2023)Kamalloo, Dziri, Clarke, and
  Rafiei}]{2023.acl-long.307}
Ehsan Kamalloo, Nouha Dziri, Charles Clarke, and Davood Rafiei. 2023.
\newblock \href {https://doi.org/10.18653/v1/2023.acl-long.307} {Evaluating
  open-domain question answering in the era of large language models}.
\newblock In \emph{Proceedings of the 61st Annual Meeting of the Association
  for Computational Linguistics (Volume 1: Long Papers)}, pages 5591--5606,
  Toronto, Canada. Association for Computational Linguistics.

\bibitem[{Kandpal et~al.(2023)Kandpal, Deng, Roberts, Wallace, and
  Raffel}]{LLMLearnLongTail?}
Nikhil Kandpal, Haikang Deng, Adam Roberts, Eric Wallace, and Colin Raffel.
  2023.
\newblock \href {https://proceedings.mlr.press/v202/kandpal23a/kandpal23a.pdf}
  {Large language models struggle to learn long-tail knowledge}.
\newblock In \emph{International Conference on Machine Learning}, pages
  15696--15707. PMLR.

\bibitem[{Karpukhin et~al.(2020)Karpukhin, Oguz, Min, Lewis, Wu, Edunov, Chen,
  and Yih}]{2020.emnlp-main.550}
Vladimir Karpukhin, Barlas Oguz, Sewon Min, Patrick Lewis, Ledell Wu, Sergey
  Edunov, Danqi Chen, and Wen-tau Yih. 2020.
\newblock \href {https://doi.org/10.18653/v1/2020.emnlp-main.550} {Dense
  passage retrieval for open-domain question answering}.
\newblock In \emph{Proceedings of the 2020 Conference on Empirical Methods in
  Natural Language Processing (EMNLP)}, pages 6769--6781, Online. Association
  for Computational Linguistics.

\bibitem[{Lewis et~al.(2020{\natexlab{a}})Lewis, Liu, Goyal, Ghazvininejad,
  Mohamed, Levy, Stoyanov, and Zettlemoyer}]{2020.acl-main.703}
Mike Lewis, Yinhan Liu, Naman Goyal, Marjan Ghazvininejad, Abdelrahman Mohamed,
  Omer Levy, Veselin Stoyanov, and Luke Zettlemoyer. 2020{\natexlab{a}}.
\newblock \href {https://doi.org/10.18653/v1/2020.acl-main.703} {{BART}:
  Denoising sequence-to-sequence pre-training for natural language generation,
  translation, and comprehension}.
\newblock In \emph{Proceedings of the 58th Annual Meeting of the Association
  for Computational Linguistics}, pages 7871--7880, Online. Association for
  Computational Linguistics.

\bibitem[{Lewis et~al.(2020{\natexlab{b}})Lewis, Perez, Piktus, Petroni,
  Karpukhin, Goyal, K{\"u}ttler, Lewis, Yih, Rockt{\"a}schel et~al.}]{RAG}
Patrick Lewis, Ethan Perez, Aleksandra Piktus, Fabio Petroni, Vladimir
  Karpukhin, Naman Goyal, Heinrich K{\"u}ttler, Mike Lewis, Wen-tau Yih, Tim
  Rockt{\"a}schel, et~al. 2020{\natexlab{b}}.
\newblock \href
  {https://proceedings.neurips.cc/paper/2020/file/6b493230205f780e1bc26945df7481e5-Paper.pdf}
  {Retrieval-augmented generation for knowledge-intensive nlp tasks}.
\newblock \emph{Advances in Neural Information Processing Systems},
  33:9459--9474.

\bibitem[{Lin et~al.(2021)Lin, Ma, Lin, Yang, Pradeep, and Nogueira}]{Pyserini}
Jimmy Lin, Xueguang Ma, Sheng-Chieh Lin, Jheng-Hong Yang, Ronak Pradeep, and
  Rodrigo Nogueira. 2021.
\newblock \href {https://dl.acm.org/doi/10.1145/3404835.3463238} {{Pyserini}: A
  {Python} toolkit for reproducible information retrieval research with sparse
  and dense representations}.
\newblock In \emph{Proceedings of the 44th Annual International ACM SIGIR
  Conference on Research and Development in Information Retrieval (SIGIR
  2021)}, pages 2356--2362.

\bibitem[{Lukovnikov et~al.(2019)Lukovnikov, Fischer, and
  Lehmann}]{arXiv:2001.11985}
Denis Lukovnikov, Asja Fischer, and Jens Lehmann. 2019.
\newblock \href {https://arxiv.org/pdf/2001.11985.pdf} {Pretrained transformers
  for simple question answering over knowledge graphs}.
\newblock In \emph{The Semantic Web--ISWC 2019: 18th International Semantic Web
  Conference, Auckland, New Zealand, October 26--30, 2019, Proceedings, Part I
  18}, pages 470--486. Springer.

\bibitem[{Mohammed et~al.(2018)Mohammed, Shi, and Lin}]{N18-2047}
Salman Mohammed, Peng Shi, and Jimmy Lin. 2018.
\newblock \href {https://doi.org/10.18653/v1/N18-2047} {Strong baselines for
  simple question answering over knowledge graphs with and without neural
  networks}.
\newblock In \emph{Proceedings of the 2018 Conference of the North {A}merican
  Chapter of the Association for Computational Linguistics: Human Language
  Technologies, Volume 2 (Short Papers)}, pages 291--296, New Orleans,
  Louisiana. Association for Computational Linguistics.

\bibitem[{Ni et~al.(2022)Ni, Qu, Lu, Dai, Hernandez~Abrego, Ma, Zhao, Luan,
  Hall, Chang, and Yang}]{2022.emnlp-main.669}
Jianmo Ni, Chen Qu, Jing Lu, Zhuyun Dai, Gustavo Hernandez~Abrego, Ji~Ma,
  Vincent Zhao, Yi~Luan, Keith Hall, Ming-Wei Chang, and Yinfei Yang. 2022.
\newblock \href {https://doi.org/10.18653/v1/2022.emnlp-main.669} {Large dual
  encoders are generalizable retrievers}.
\newblock In \emph{Proceedings of the 2022 Conference on Empirical Methods in
  Natural Language Processing}, pages 9844--9855, Abu Dhabi, United Arab
  Emirates. Association for Computational Linguistics.

\bibitem[{Ong et~al.(2024)Ong, Shavarani, and Sarkar}]{2024.naacl-main.???}
Nicolas Ong, Hassan~S. Shavarani, and Anoop Sarkar. 2024.
\newblock \href {https://openreview.net/forum?id=p8U0sZWOrt} {Unified
  examination of entity linking in absence of candidate sets}.
\newblock In \emph{Proceedings of the 2024 Conference of the North American
  Chapter of the Association for Computational Linguistics: Human Language
  Technologies}, Mexico. Association for Computational Linguistics.

\bibitem[{OpenAI(2023)}]{GPT4}
OpenAI. 2023.
\newblock \href {http://arxiv.org/abs/2303.08774} {Gpt-4 technical report}.

\bibitem[{Peng et~al.(2023)Peng, Galley, He, Cheng, Xie, Hu, Huang, Liden, Yu,
  Chen et~al.}]{arXiv:2302.12813}
Baolin Peng, Michel Galley, Pengcheng He, Hao Cheng, Yujia Xie, Yu~Hu, Qiuyuan
  Huang, Lars Liden, Zhou Yu, Weizhu Chen, et~al. 2023.
\newblock \href {https://arxiv.org/pdf/2302.12813.pdf} {Check your facts and
  try again: Improving large language models with external knowledge and
  automated feedback}.
\newblock \emph{arXiv preprint arXiv:2302.12813}.

\bibitem[{Raffel et~al.(2020)Raffel, Shazeer, Roberts, Lee, Narang, Matena,
  Zhou, Li, and Liu}]{T5}
Colin Raffel, Noam Shazeer, Adam Roberts, Katherine Lee, Sharan Narang, Michael
  Matena, Yanqi Zhou, Wei Li, and Peter~J Liu. 2020.
\newblock \href {https://www.jmlr.org/papers/volume21/20-074/20-074.pdf}
  {Exploring the limits of transfer learning with a unified text-to-text
  transformer}.
\newblock \emph{The Journal of Machine Learning Research}, 21(1):5485--5551.

\bibitem[{Rajpurkar et~al.(2016)Rajpurkar, Zhang, Lopyrev, and
  Liang}]{D16-1264}
Pranav Rajpurkar, Jian Zhang, Konstantin Lopyrev, and Percy Liang. 2016.
\newblock \href {https://doi.org/10.18653/v1/D16-1264} {{SQ}u{AD}: 100,000+
  questions for machine comprehension of text}.
\newblock In \emph{Proceedings of the 2016 Conference on Empirical Methods in
  Natural Language Processing}, pages 2383--2392, Austin, Texas. Association
  for Computational Linguistics.

\bibitem[{Ram et~al.(2023)Ram, Levine, Dalmedigos, Muhlgay, Shashua,
  Leyton-Brown, and Shoham}]{2023.tacl-1.75}
Ori Ram, Yoav Levine, Itay Dalmedigos, Dor Muhlgay, Amnon Shashua, Kevin
  Leyton-Brown, and Yoav Shoham. 2023.
\newblock \href {https://doi.org/10.1162/tacl_a_00605} {In-context
  retrieval-augmented language models}.
\newblock \emph{Transactions of the Association for Computational Linguistics},
  11:1316--1331.

\bibitem[{Ranjan and Balabantaray(2016)}]{IEEE_7917964}
Prakash Ranjan and Rakesh~Chandra Balabantaray. 2016.
\newblock \href
  {https://ieeexplore.ieee.org/stamp/stamp.jsp?tp=&arnumber=7917964} {Question
  answering system for factoid based question}.
\newblock In \emph{2016 2nd International Conference on Contemporary Computing
  and Informatics (IC3I)}, pages 221--224. IEEE.

\bibitem[{Roberts et~al.(2020)Roberts, Raffel, and
  Shazeer}]{2020.emnlp-main.437}
Adam Roberts, Colin Raffel, and Noam Shazeer. 2020.
\newblock \href {https://doi.org/10.18653/v1/2020.emnlp-main.437} {How much
  knowledge can you pack into the parameters of a language model?}
\newblock In \emph{Proceedings of the 2020 Conference on Empirical Methods in
  Natural Language Processing (EMNLP)}, pages 5418--5426, Online. Association
  for Computational Linguistics.

\bibitem[{Robertson et~al.(2009)Robertson, Zaragoza et~al.}]{BM25}
Stephen Robertson, Hugo Zaragoza, et~al. 2009.
\newblock \href {https://www.nowpublishers.com/article/Details/INR-019} {The
  probabilistic relevance framework: Bm25 and beyond}.
\newblock \emph{Foundations and Trends{\textregistered} in Information
  Retrieval}, 3(4):333--389.

\bibitem[{Robertson et~al.(1994)Robertson, Walker, Jones, Hancock-Beaulieu, and
  Gatford}]{BM25Original}
Stephen~E. Robertson, Steve Walker, Susan Jones, Micheline Hancock-Beaulieu,
  and Mike Gatford. 1994.
\newblock \href {https://api.semanticscholar.org/CorpusID:3946054} {Okapi at
  trec-3}.
\newblock In \emph{Text Retrieval Conference}.

\bibitem[{Salnikov et~al.(2023)Salnikov, Le, Rajput, Nikishina, Braslavski,
  Malykh, and Panchenko}]{2023.paclic-1.63}
Mikhail Salnikov, Hai Le, Prateek Rajput, Irina Nikishina, Pavel Braslavski,
  Valentin Malykh, and Alexander Panchenko. 2023.
\newblock \href {https://aclanthology.org/2023.paclic-1.63} {Large language
  models meet knowledge graphs to answer factoid questions}.
\newblock In \emph{Proceedings of the 37th Pacific Asia Conference on Language,
  Information and Computation}, pages 635--644, Hong Kong, China. Association
  for Computational Linguistics.

\bibitem[{Santhanam et~al.(2022)Santhanam, Khattab, Saad-Falcon, Potts, and
  Zaharia}]{2022.naacl-main.272}
Keshav Santhanam, Omar Khattab, Jon Saad-Falcon, Christopher Potts, and Matei
  Zaharia. 2022.
\newblock \href {https://doi.org/10.18653/v1/2022.naacl-main.272}
  {{C}ol{BERT}v2: Effective and efficient retrieval via lightweight late
  interaction}.
\newblock In \emph{Proceedings of the 2022 Conference of the North American
  Chapter of the Association for Computational Linguistics: Human Language
  Technologies}, pages 3715--3734, Seattle, United States. Association for
  Computational Linguistics.

\bibitem[{Sciavolino et~al.(2021)Sciavolino, Zhong, Lee, and
  Chen}]{2021.emnlp-main.496}
Christopher Sciavolino, Zexuan Zhong, Jinhyuk Lee, and Danqi Chen. 2021.
\newblock \href {https://doi.org/10.18653/v1/2021.emnlp-main.496} {Simple
  entity-centric questions challenge dense retrievers}.
\newblock In \emph{Proceedings of the 2021 Conference on Empirical Methods in
  Natural Language Processing}, pages 6138--6148, Online and Punta Cana,
  Dominican Republic. Association for Computational Linguistics.

\bibitem[{Shavarani and Sarkar(2023)}]{2023.emnlp-main.686}
Hassan~S. Shavarani and Anoop Sarkar. 2023.
\newblock \href {https://aclanthology.org/2023.emnlp-main.686} {{S}p{EL}:
  Structured prediction for entity linking}.
\newblock In \emph{Proceedings of the 2023 Conference on Empirical Methods in
  Natural Language Processing}, pages 11123--11137, Singapore. Association for
  Computational Linguistics.

\bibitem[{Shavarani and Sekine(2020)}]{2020.lrec-1.150}
Hassan~S. Shavarani and Satoshi Sekine. 2020.
\newblock \href {https://aclanthology.org/2020.lrec-1.150} {Multi-class
  multilingual classification of {W}ikipedia articles using extended named
  entity tag set}.
\newblock In \emph{Proceedings of the Twelfth Language Resources and Evaluation
  Conference}, pages 1197--1201, Marseille, France. European Language Resources
  Association.

\bibitem[{Shi et~al.(2023)Shi, Min, Yasunaga, Seo, James, Lewis, Zettlemoyer,
  and Yih}]{arXiv:2301.12652}
Weijia Shi, Sewon Min, Michihiro Yasunaga, Minjoon Seo, Rich James, Mike Lewis,
  Luke Zettlemoyer, and Wen-tau Yih. 2023.
\newblock \href {https://arxiv.org/pdf/2301.12652.pdf} {Replug:
  Retrieval-augmented black-box language models}.
\newblock \emph{arXiv preprint arXiv:2301.12652}.

\bibitem[{Shuster et~al.(2021)Shuster, Poff, Chen, Kiela, and
  Weston}]{2021.findings-emnlp.320}
Kurt Shuster, Spencer Poff, Moya Chen, Douwe Kiela, and Jason Weston. 2021.
\newblock \href {https://doi.org/10.18653/v1/2021.findings-emnlp.320}
  {Retrieval augmentation reduces hallucination in conversation}.
\newblock In \emph{Findings of the Association for Computational Linguistics:
  EMNLP 2021}, pages 3784--3803, Punta Cana, Dominican Republic. Association
  for Computational Linguistics.

\bibitem[{Singh et~al.(2021)Singh, Reddy, Hamilton, Dyer, and Yogatama}]{EMDR2}
Devendra Singh, Siva Reddy, Will Hamilton, Chris Dyer, and Dani Yogatama. 2021.
\newblock \href
  {https://proceedings.neurips.cc/paper_files/paper/2021/file/da3fde159d754a2555eaa198d2d105b2-Paper.pdf}
  {End-to-end training of multi-document reader and retriever for open-domain
  question answering}.
\newblock \emph{Advances in Neural Information Processing Systems},
  34:25968--25981.

\bibitem[{Smith et~al.(2008)Smith, Heilman, and Hwa}]{FactoidQA}
Noah~A Smith, Michael Heilman, and Rebecca Hwa. 2008.
\newblock \href
  {https://www.cs.cmu.edu/~nasmith/papers/smith+heilman+hwa.nsf08.pdf}
  {Question generation as a competitive undergraduate course project}.
\newblock In \emph{Proceedings of the NSF Workshop on the Question Generation
  Shared Task and Evaluation Challenge}, volume~9.

\bibitem[{Sun et~al.(2018)Sun, Dhingra, Zaheer, Mazaitis, Salakhutdinov, and
  Cohen}]{D18-1455}
Haitian Sun, Bhuwan Dhingra, Manzil Zaheer, Kathryn Mazaitis, Ruslan
  Salakhutdinov, and William Cohen. 2018.
\newblock \href {https://doi.org/10.18653/v1/D18-1455} {Open domain question
  answering using early fusion of knowledge bases and text}.
\newblock In \emph{Proceedings of the 2018 Conference on Empirical Methods in
  Natural Language Processing}, pages 4231--4242, Brussels, Belgium.
  Association for Computational Linguistics.

\bibitem[{Voorhees and Harman(1999)}]{MRR}
Ellen~M Voorhees and Donna Harman. 1999.
\newblock Overview of the eighth text retrieval conference (trec-8).
\newblock In \emph{Proceedings of the Eighth Text REtrieval Conference
  (TREC-8), NIST Special Publication}.

\bibitem[{Xiong et~al.(2021)Xiong, Xiong, Li, Tang, Liu, Bennett, Ahmed, and
  Overwijk}]{ANCE}
Lee Xiong, Chenyan Xiong, Ye~Li, Kwok-Fung Tang, Jialin Liu, Paul~N. Bennett,
  Junaid Ahmed, and Arnold Overwijk. 2021.
\newblock \href {https://openreview.net/forum?id=zeFrfgyZln} {Approximate
  nearest neighbor negative contrastive learning for dense text retrieval}.
\newblock In \emph{International Conference on Learning Representations}.

\bibitem[{Yamada et~al.(2021)Yamada, Asai, and Hajishirzi}]{2021.acl-short.123}
Ikuya Yamada, Akari Asai, and Hannaneh Hajishirzi. 2021.
\newblock \href {https://doi.org/10.18653/v1/2021.acl-short.123} {Efficient
  passage retrieval with hashing for open-domain question answering}.
\newblock In \emph{Proceedings of the 59th Annual Meeting of the Association
  for Computational Linguistics and the 11th International Joint Conference on
  Natural Language Processing (Volume 2: Short Papers)}, pages 979--986,
  Online. Association for Computational Linguistics.

\bibitem[{Yu et~al.(2023)Yu, Zhang, Liang, Jiang, and
  Sabharwal}]{arXiv:2305.14002}
Wenhao Yu, Zhihan Zhang, Zhenwen Liang, Meng Jiang, and Ashish Sabharwal. 2023.
\newblock \href {https://arxiv.org/pdf/2305.14002.pdf} {Improving language
  models via plug-and-play retrieval feedback}.
\newblock \emph{arXiv preprint arXiv:2305.14002}.

\bibitem[{Zhan et~al.(2020{\natexlab{a}})Zhan, Mao, Liu, Zhang, and Ma}]{LTRe}
Jingtao Zhan, Jiaxin Mao, Yiqun Liu, Min Zhang, and Shaoping Ma.
  2020{\natexlab{a}}.
\newblock \href {https://arxiv.org/pdf/2010.10469} {Learning to retrieve: How
  to train a dense retrieval model effectively and efficiently}.
\newblock \emph{arXiv preprint arXiv:2010.10469}.

\bibitem[{Zhan et~al.(2020{\natexlab{b}})Zhan, Mao, Liu, Zhang, and
  Ma}]{RepBERT}
Jingtao Zhan, Jiaxin Mao, Yiqun Liu, Min Zhang, and Shaoping Ma.
  2020{\natexlab{b}}.
\newblock \href {https://arxiv.org/pdf/2006.15498} {Repbert: Contextualized
  text embeddings for first-stage retrieval}.
\newblock \emph{arXiv preprint arXiv:2006.15498}.

\bibitem[{Zhang et~al.(2023)Zhang, Chen, Xu, Cao, Chen, Cohn, and
  Fang}]{2023.acl-long.808}
Qin Zhang, Shangsi Chen, Dongkuan Xu, Qingqing Cao, Xiaojun Chen, Trevor Cohn,
  and Meng Fang. 2023.
\newblock \href {https://doi.org/10.18653/v1/2023.acl-long.808} {A survey for
  efficient open domain question answering}.
\newblock In \emph{Proceedings of the 61st Annual Meeting of the Association
  for Computational Linguistics (Volume 1: Long Papers)}, pages 14447--14465,
  Toronto, Canada. Association for Computational Linguistics.

\bibitem[{Zhu et~al.(2021)Zhu, Lei, Wang, Zheng, Poria, and
  Chua}]{arXiv:2101.00774}
Fengbin Zhu, Wenqiang Lei, Chao Wang, Jianming Zheng, Soujanya Poria, and
  Tat-Seng Chua. 2021.
\newblock \href {https://arxiv.org/pdf/2101.00774.pdf} {Retrieving and reading:
  A comprehensive survey on open-domain question answering}.
\newblock \emph{arXiv preprint arXiv:2101.00774}.

\end{thebibliography}
\appendix
\section{Example Prompts for Different Experimental Settings}\label{appendix_a}
In this section, we present the prompts used in our experimental settings. For each setting, we provide the prompt template, and explain the processes needed to obtain the augmentation documents if a Retrieval-Augmented setting is being discussed.

\subsection{Closed-book Setting}\label{sec:appendix_cb_setting}
In this setting, we do not have any augmentation documents, so the prompt contains the instruction, followed by the question:\\

\begin{tikzpicture}
    \node[draw, rounded corners, fill=gray!20, text width=0.85\columnwidth, inner sep=10pt] {
    \vspace{-\baselineskip}
        \begin{Verbatim}
Answer this question:
Q: {question}
A:
        \end{Verbatim}
    };
\end{tikzpicture}

Here is an example prompt with the question mentioned in Figure \ref{fig:dpr_to_entity_retrieval_comparison} and Table \ref{tab:raqa_comparison_examples}:\\

\begin{tikzpicture}
    \node[draw, rounded corners, fill=gray!20, text width=0.85\columnwidth, inner sep=10pt] {
    \vspace{-\baselineskip}
        \begin{Verbatim}
Answer this question:
Q: What is the capital of Seine-Saint
   -Denis?
A:
        \end{Verbatim}
    };
\end{tikzpicture}

\subsection{Retrieval-Augmented Settings}\label{sec:appendix_ra_setting}
In this setting, we examine two variations of prompts based on the number of available augmented documents. For a single document, the prompt is as follows:\\

\begin{tikzpicture}
    \node[draw, rounded corners, fill=gray!20, text width=0.85\columnwidth, inner sep=10pt] {
    \vspace{-\baselineskip}
        \begin{Verbatim}
{document}

Based on this text, answer this 
question:
Q: {question}
A:
        \end{Verbatim}
    };
\end{tikzpicture}

When multiple documents are available, they are presented sequentially, followed by the instruction and question:\\

\begin{tikzpicture}
    \node[draw, rounded corners, fill=gray!20, text width=0.85\columnwidth, inner sep=10pt] {
    \vspace{-\baselineskip}
        \begin{Verbatim}
{document1}

{document2}

...

{documentN}

Based on these texts, answer 
this question:
Q: {question}
A:
        \end{Verbatim}
    };
\end{tikzpicture}

\begin{table*}[!ht]
    \centering
    \begin{tabular}{l|p{0.9\linewidth}}
        \toprule    
        Doc\# & Content \\
        \midrule
        1 & Pierrefitte-sur-Seine<newline>Pierrefitte-sur-Seine Pierrefitte-sur-Seine is a commune in the Seine-Saint-Denis department and Ile-de-France region of France. Today forming part of the northern suburbs of Paris, Pierrefitte lies from the centre of the French capital. The town is served by Pierrefitte - Stains railway station on line D of the RER regional suburban rail network. The south of the commune, where the National Archives of France relocated in 2013, is also served by Saint-Denis - Universite station on Paris Metro Line 13. This station lies on the border between the communes of Pierrefitte-sur-Seine and Saint-Denis. Primary and secondary schools in the commune include: \\\midrule
        2 & "Saint-Ouen, Seine-Saint-Denis"<newline>Saint-Ouen, Seine-Saint-Denis Saint-Ouen () is a commune in the Seine-Saint-Denis department. It is located in the northern suburbs of Paris, France, from the centre of Paris. The communes neighbouring Saint-Ouen are Paris, to the south, Clichy, to the west, Asnieres-sur-Seine and L'Ile-Saint-Denis, to the north, and Saint-Denis to the east. The commune of Saint-Ouen is part of the canton of Saint-Ouen, which also includes L'Ile-Saint-Denis and part of Epinay-sur-Seine. Saint-Ouen also includes the Cimetiere de Saint-Ouen. On 1 January 1860, the city of Paris was enlarged by annexing neighbouring communes. On that occasion, a part of the commune of Saint-Ouen\\\midrule
        3 & "Ile-de-France"<newline>of France. The population of immigrants is more widely distributed throughout the region than it was in the early 2000s, though the concentrations remain high in certain areas, particularly Paris and the department of Seine-Saint-Denis. The proportion of residents born outside of Metropolitan France has dropped since the 1999 census (19.7 percent) and the 2010 census (23 percent). . The Petite Couronne (Little Crown, i.e. ""Inner Ring"") is formed by the 3 departments of Ile-de-France bordering with the French capital and forming a geographical ""crown"" around it. The departments, until 1968 part of the disbanded Seine department, are Hauts-de-Seine, Seine-Saint-Denis \\\midrule
        4 & "Saint-Denis, Seine-Saint-Denis"<newline>Saint-Denis, Seine-Saint-Denis Saint-Denis () is a commune in the northern suburbs of Paris, France. It is located from the centre of Paris. Saint-Denis is a subprefecture () of the department of Seine-Saint-Denis, being the seat of the arrondissement of Saint-Denis. Saint-Denis is home to the royal necropolis of the Basilica of Saint Denis and was also the location of the associated abbey. It is also home to France's national football and rugby stadium, the Stade de France, built for the 1998 FIFA World Cup. Saint-Denis is a formerly industrial suburb currently changing its economic base. Inhabitants of Saint-Denis are called \\
        \bottomrule
    \end{tabular}
    \caption{Top 4 documents retrieved from the BM25 Lucene index for the question \texttt{What is the capital of Seine-Saint-Denis?} from the EntityQuestions (dev) dataset.}
    \label{tab:bm25_for_ssd}
\end{table*}
\begin{table*}[!ht]
    \centering
    \begin{tabular}{l|p{0.9\linewidth}}
        \toprule    
        Doc\# & Content \\
        \midrule
         1 &  "L'Ile-Saint-Denis"<newline>L'Ile-Saint-Denis L'Ile-Saint-Denis (the island of Saint Denis) is a commune in the northern suburbs of Paris, France. It is located from the center of Paris. The commune is entirely contained on an island of the Seine River, hence its name. Several transit connections are located nearby. The closest station to L'Ile-Saint-Denis is Saint-Denis station, which is an interchange station on Paris RER line D and on the Transilien Paris - Nord suburban rail line. This station is located in the neighboring commune of Saint-Denis, from the town center of L'Ile-Saint-Denis. Tram T1 stops near Ile-Saint-Denis's town hall. Bus route 237\\\midrule
         2 &  "15th arrondissement of Paris"<newline>15th arrondissement of Paris The 15th arrondissement of Paris (""XV arrondissement"") is one of the 20 arrondissements of the capital city of France. In spoken French, this arrondissement is referred to as ""quinzieme"". The arrondissement, called Vaugirard, is situated on the left bank of the River Seine. Sharing the Montparnasse district with the 6th and 14th arrondissements, it is the city's most populous arrondissement. The ""Tour Montparnasse"" - the tallest skyscraper in Paris - and the neighbouring Gare Montparnasse are both located in the 15th arrondissement, at its border with the 14th. It is also home to the convention center\\\midrule
         3 &  "L'Ile-Saint-Denis"<newline>few of the students were White. There are three primary schools in the commune: Ecole Samira Bellil, Ecole Paul Langevin, and Ecole Jean Lurcat. College Alfred Sisley, a junior high school, is on the island. L'Ile-Saint-Denis L'Ile-Saint-Denis (the island of Saint Denis) is a commune in the northern suburbs of Paris, France. It is located from the center of Paris. The commune is entirely contained on an island of the Seine River, hence its name. Several transit connections are located nearby. The closest station to L'Ile-Saint-Denis is Saint-Denis station, which is an interchange station on Paris RER line D and\\\midrule
         4 &  "2nd arrondissement of Paris"<newline>2nd arrondissement of Paris The 2nd arrondissement of Paris (""II arrondissement"") is one of the 20 arrondissements of the capital city of France. In spoken French, this arrondissement is colloquially referred to as ""deuxieme"" (second/the second). Also known as Bourse, this arrondissement is located on the right bank of the River Seine. The 2nd arrondissement, together with the adjacent 8th and 9th arrondissements, hosts an important business district, centred on the Paris Opera, which houses the city's most dense concentration of business activities. The arrondissement contains the former Paris Bourse (stock exchange) and a large number of banking headquarters, as\\
         \bottomrule
    \end{tabular}
    \caption{Top 4 documents retrieved from the DPR index for the question \texttt{What is the capital of Seine-Saint-Denis?} from the EntityQuestions (dev) dataset.}
    \label{tab:dpr_for_ssd}
\end{table*}
\begin{table*}[!ht]
    \centering
    \begin{tabular}{l|p{0.9\linewidth}}
        \toprule    
        Doc\# & Content \\
        \midrule
         1 &  "L'Ile-Saint-Denis"<newline>L'Ile-Saint-Denis L'Ile-Saint-Denis (the island of Saint Denis) is a commune in the northern suburbs of Paris, France. It is located from the center of Paris. The commune is entirely contained on an island of the Seine River, hence its name. Several transit connections are located nearby. The closest station to L'Ile-Saint-Denis is Saint-Denis station, which is an interchange station on Paris RER line D and on the Transilien Paris - Nord suburban rail line. This station is located in the neighboring commune of Saint-Denis, from the town center of L'Ile-Saint-Denis. Tram T1 stops near Ile-Saint-Denis's town hall. Bus route 237\\\midrule
         2 &  "L'Ile-Saint-Denis"<newline>few of the students were White. There are three primary schools in the commune: Ecole Samira Bellil, Ecole Paul Langevin, and Ecole Jean Lurcat. College Alfred Sisley, a junior high school, is on the island. L'Ile-Saint-Denis L'Ile-Saint-Denis (the island of Saint Denis) is a commune in the northern suburbs of Paris, France. It is located from the center of Paris. The commune is entirely contained on an island of the Seine River, hence its name. Several transit connections are located nearby. The closest station to L'Ile-Saint-Denis is Saint-Denis station, which is an interchange station on Paris RER line D and\\\midrule
         3 &  "Saint-Denis, Seine-Saint-Denis"<newline>one private elementary, middle, and high school (""Ensemble Scolaire Jean-Baptiste de la Salle-Notre Dame de la Compassion"") and one private middle and high school (""College et lycee Saint-Vincent-de-Paul""). Saint-Denis is twinned with: Saint-Denis, Seine-Saint-Denis Saint-Denis () is a commune in the northern suburbs of Paris, France. It is located from the centre of Paris. Saint-Denis is a subprefecture () of the department of Seine-Saint-Denis, being the seat of the arrondissement of Saint-Denis. Saint-Denis is home to the royal necropolis of the Basilica of Saint Denis and was also the location of the associated abbey. It is also home to France's\\\midrule
         4 &  "Saint-Ouen, Seine-Saint-Denis"<newline>Saint-Ouen, Seine-Saint-Denis Saint-Ouen () is a commune in the Seine-Saint-Denis department. It is located in the northern suburbs of Paris, France, from the centre of Paris. The communes neighbouring Saint-Ouen are Paris, to the south, Clichy, to the west, Asnieres-sur-Seine and L'Ile-Saint-Denis, to the north, and Saint-Denis to the east. The commune of Saint-Ouen is part of the canton of Saint-Ouen, which also includes L'Ile-Saint-Denis and part of Epinay-sur-Seine. Saint-Ouen also includes the Cimetiere de Saint-Ouen. On 1 January 1860, the city of Paris was enlarged by annexing neighbouring communes. On that occasion, a part of the commune of Saint-Ouen\\
         \bottomrule
    \end{tabular}
    \caption{Top 4 documents retrieved from the ANCE index for the question \texttt{What is the capital of Seine-Saint-Denis?} from the EntityQuestions (dev) dataset.}
    \label{tab:ance_for_ssd}
\end{table*}

Next, we examine the various Retrieval-Augmentation techniques studied in this paper: BM25, DPR, and ANCE, showcasing their top four retrieved documents for \texttt{What is the capital of Seine-Saint-Denis?}. Tables \ref{tab:bm25_for_ssd}, \ref{tab:dpr_for_ssd}, and \ref{tab:ance_for_ssd} present these retrieved documents. The finalized prompt template will include the four retrieved documents alongside the question, as previously discussed.

In analyzing the retrieved documents, you can verify the originating Wikipedia articles mentioned in the beginning of each passage. Notably, passages are drawn from three or four different articles, and given the entity-centric nature of the question, relying on multiple sources could mislead the LLM, as suggested by \citet{arXiv:2401.14887}. Additionally, these methods primarily focus on lexical similarity, particularly the presence of \texttt{capital}, \texttt{Seine}, \texttt{Saint}, and \texttt{Denis}. However, this focus has not consistently led to retrieval of passages containing the correct answer: \texttt{Bobigny}.

\subsection{\textit{Entity Retrieval} Settings}

For \textit{Entity Retrieval}, we utilize an entity linker to identify entities within the question. In this study, we employed \textsc{SpEL}, though any suitable entity linking method can be used. The primary requirement is that the linker accepts a string (the question) as input and returns a list containing (\texttt{begin character}, \texttt{end character}, \texttt{identified entity}) tuples. The \texttt{begin character} and \texttt{end character} values help determine the order of entity annotations in the text, ensuring proper sequence if the returned list is unordered.

The \texttt{identified entity} values are then used to search the Wikipedia dump on disk, fetching articles corresponding to the identified entities. Efficient implementation of this lookup process is crucial for the method's performance. Our approach involves a two-step indexing and lookup process. First, we index the file bytes marking the beginning of each Wikipedia article in the dump file. When an article is needed, we use this index to locate the byte number and employ the \texttt{seek} method to navigate to the correct position in the file and read the article.

After gathering the relevant articles, we truncate each one to the first $W$ words (suffixed with the Wikipedia identifier, as per convention) to create a list of augmentation documents to accompany the question when querying the LLM.

To prompt the LLM, we use the same prompts previously mentioned. If no entities are found in the question, we refer to the prompt in Appendix \ref{sec:appendix_cb_setting}. If one entity is recognized, resulting in one augmentation document, we use the first prompt from Appendix \ref{sec:appendix_ra_setting}. If multiple entities are identified, we use the second prompt from the same appendix section. In rare cases where the number of identified entities exceeds $k$ (the expected number of documents to retrieve), we simply consider the first $k$ unique entities to form the list of augmentation documents.

\begin{table*}[!ht]
    \centering
    \begin{tabular}{l|p{0.9\linewidth}}
        \toprule    
        Doc\# & Content \\
        \midrule
         1 &  Seine-Saint-Denis<newline>Seine-Saint-Denis In 2019, it had a population of 1,644,903 across 40 communes. In French, the learned but rarely used demonym for the inhabitants of Seine-Saint-Denis is ; more common is . The department is surrounded by the departments of Hauts-de-Seine, Val-de-Marne, Paris, Val-d'Oise, and Seine-et-Marne. It is thus the only one of the five French departments surrounded entirely by other departments of the same region. Image:Petite couronne.png The most populous commune is Saint-Denis; the prefecture \setlength{\fboxsep}{1pt}\colorbox{pink}{Bobigny} is the eleventh-most populous. As of 2019, there are 5 communes with more than 70,000 inhabitants: is made up of three departmental and 40\\
         \bottomrule
    \end{tabular}
    \caption{The only document retrieved by \textit{Entity Retrieval} using \textsc{SpEL} annotations for the question \texttt{What is the capital of Seine-Saint-Denis?} from the EntityQuestions (dev) dataset. \textsc{SpEL} identifies only one entity in the question: \texttt{Seine-Saint-Denis} and returns the first 100 words (considering $W$=100) of its Wikipedia article as the retrieved document. The answer to the question: \texttt{Bobigny} is highlighted for ease of verification.}
    \label{tab:er_for_ssd}
\end{table*}

Table \ref{tab:er_for_ssd} presents the single document retrieved for \texttt{What is the capital of Seine-Saint-Denis?}, which contains the answer: \texttt{Bobigny}. Examining the lexical distribution in this document, we observe that unlike the BM25 method, \textit{Entity Retrieval} treats the salient entity \texttt{Seine-Saint-Denis} as an atomic term rather than emphasizing each word in the question. This focused approach, coupled with the retrieval of fewer documents, allows the model to concentrate on the relevant information, reducing noise and potential confusion.

However, the effectiveness of \textit{Entity Retrieval} in real-world scenarios, where question entity annotations are not available, largely depends on the quality of the entity linker used to identify salient entities in the question. Therefore, further research into developing more accurate entity linking models could enhance \textit{Entity Retrieval} performance.

\end{document}